\documentclass[runningheads]{llncs}
\usepackage[T1]{fontenc}
\usepackage[braket, qm]{qcircuit}
\usepackage{pifont}
\usepackage{bbding}
\usepackage{hyperref}
\usepackage{breakurl}
\usepackage{amsmath,amssymb,amsfonts}
\usepackage{algorithmic}
\usepackage{subfigure}
\usepackage{qcircuit}
\usepackage{booktabs}
\usepackage{array}

\usepackage{graphicx}
\usepackage{xcolor}

\definecolor{color_ripple_carry}{HTML}{EA4335} 
\definecolor{color_carry_look_ahead}{HTML}{4285F4} 
\definecolor{color_hybrid}{HTML}{34A853} 
\definecolor{color_carry_save}{HTML}{F97D53} 

\definecolor{color_restoring}{HTML}{3C93C2} 
\definecolor{color_non_restoring}{HTML}{226E9C} 
\definecolor{color_long_division}{HTML}{0D4A70} 
\definecolor{color_goldschmidt}{HTML}{EA4335} 
\definecolor{color_newton_raphson}{HTML}{F97B4F} 

\definecolor{color_schoolbook}{HTML}{AF58BA} 
\definecolor{color_shift_and_add}{HTML}{FF571F} 
\definecolor{color_karatsuba}{HTML}{EA4335} 
\definecolor{color_toom_cook}{HTML}{4285F4} 
\definecolor{color_windowed}{HTML}{FFC61E} 
\definecolor{color_schonhage_strassen}{HTML}{7065BD} 
\definecolor{color_wallace_tree}{HTML}{34A853} 

\definecolor{color_toom_cook_2}{HTML}{46AEA0} 
\definecolor{color_toom_cook_2_5}{HTML}{089099} 
\definecolor{color_toom_cook_3}{HTML}{0071BB} 
\definecolor{color_toom_cook_4}{HTML}{045275} 
\definecolor{color_toom_cook_8}{HTML}{003147} 

\definecolor{color_Mod_Exp}{HTML}{4285F4} 
\definecolor{color_qft}{HTML}{EA4335} 
\definecolor{color_point_add}{HTML}{34A853} 

\begin{document}
%
\title{Quantum Arithmetic Circuits in Public-Key Cryptography}
%
%
\author{Siyi Wang\inst{1}\thanks{Corresponding author. Email: \href{mailto:siyi002@e.ntu.edu.sg}{\tt siyi002@e.ntu.edu.sg}.} \and
Kyungbae Jang \inst{2}\and
Hyunji Kim \inst{2}\and
Anik Basu Bhaumik\inst{1} \and
Anubhab Baksi\inst{3}\and 
Hwajeong Seo \inst{2}\and
Anupam Chattopadhyay\inst{1}}
\authorrunning{Wang et al.}
%
\institute{Nanyang Technological University, Singapore, Singapore \and
Hansung University, Seoul, South Korea \and
Lunds Universitets, Lund, Sweden 
}

\maketitle              
\begin{abstract}
Quantum computing has advanced rapidly in recent decades, driven by developments across the technology stack, including quantum error-correcting codes and efficient quantum algorithms. Among these, quantum arithmetic circuits serve as fundamental building blocks for various promising algorithms. Despite their crucial role, the design of quantum arithmetic circuits faces challenges arising from the no-cloning theorem, qubit limitations, and circuit depth constraints, which significantly impact the efficiency of large-scale quantum computing.
We provide an overview of quantum arithmetic circuits in the context of public-key cryptanalysis, with particular emphasis on optimization strategies such as measurement-based uncomputation and conditionally clean ancilla. We review state-of-the-art designs for essential arithmetic operations in public-key cryptanalysis such as addition, multiplication, and modular exponentiation. We also present an overview of the techniques used for fault-tolerant runtime and resource estimation in quantum cryptanalysis. 
%
In brief, this chapter emphasizes strategies for designing resource-efficient quantum arithmetic circuits, providing a basis for realistic evaluations of quantum cryptanalytic capabilities.




\end{abstract}

\section{Introduction}
By leveraging the principles of quantum mechanics, quantum computing introduces new ways of processing information, with the potential to transform areas such as cryptography and materials science. Breakthroughs in the 1990s, particularly Shor’s factoring algorithm~\cite{Shor} and Grover’s search~\cite{grover1996fast}, demonstrated that quantum computers can achieve exponential and quadratic speedups over classical computing for certain problems, especially in cryptanalysis.

One of the most directly impacted areas is public-key cryptography, which underpins secure communication in modern information systems. Schemes such as RSA~\cite{rivest1978method} and elliptic curve cryptography (ECC) derive their security from the hardness of problems like integer factorization and the elliptic curve discrete logarithm problem. However, Shor’s algorithm can efficiently solve both problems on a quantum computer, rendering these widely used public-key systems vulnerable.

Specifically, for the Integer Factorization Problem (IFP) underlying RSA encryption, the most computationally demanding component of Shor’s algorithm is the modular exponentiation. This operation consists of a sequence of modular arithmetic subroutines, such as modular addition and modular multiplication.
In contrast, for the Elliptic Curve Discrete Logarithm Problem (ECDLP), which underpins ECC, Shor’s algorithm repeatedly executes point additions over a finite field. The computational complexity is mainly determined by the finite-field arithmetic operations, including modular addition, modular multiplication, and particularly modular division, the latter being the most resource-intensive operation.
The efficiency of these basic quantum arithmetic blocks directly affects the scalability and practicality of quantum cryptanalysis. Therefore, optimizing quantum arithmetic circuits has become a critical research focus~\cite{wang2025innovative,wang2025comprehensive}, as improvements in these designs can substantially reduce the resource requirements of large-scale quantum algorithms such as Shor's algorithm.

\subsection{Quantum Computing Basics}
\begin{enumerate}
    \item \textbf{Quantum bits (Qubits). }
    It is the basic unit in quantum computing, similar to classical bit but affected by quantum mechanics such as superposition and entanglement. 
    %
    Generally, qubits can be categorized into physical or logical types:
    \textit{Physical qubits}~\cite{viola2001constructing} are realized in actual hardware using techniques such as photons, trapped ions, and superconductors. However, those are prone to errors from decoherence and noise. To address this problem, \textit{logical qubits}~\cite{shaw2008encoding} are constructed from multiple physical qubits using quantum error correction, providing error-resilient abstractions that enable reliable computation.
    %
    In brief, physical qubits are realized in the actual hardware devices, while logical qubits provide abstract, stable, and error-corrected units for executing quantum algorithms.

    \item \textbf{Basic Logical Gates.} This is another basic building blocks of quantum computing. Several commonly used quantum gates, including the Hadamard, Pauli-X, Pauli-Y, Pauli-Z, and CNOT gates, are illustrated in Table ~\ref{Basic_Gates}. 
    In fault-tolerant quantum computing, the T gate is particularly costly, as it requires resource-intensive magic state distillation~\cite{bravyi2005universal}. Thus, the T gate and the Toffoli gate, which can be decomposed into multiple T gates, are among the most expensive quantum operations. 
\end{enumerate}

\vspace{-0.5cm}
\begin{table}[!ht]
\caption{Basic Quantum Logic Gates.\label{Basic_Gates}}
\setlength{\abovecaptionskip}{0cm} 
\setlength{\belowcaptionskip}{-2cm} 
\centering
\begin{tabular}{c|c|c|c|c|c|c|c|c|c}
\hline
{Operator} & Pauli-X & Pauli-Y & Pauli-Z & Hadamard & S & T & CNOT & CZ  & Toffoli \\ \hline
{Gate}     & \Qcircuit @C=1em @R=.7em {
& \gate{X} & \qw\\&{\text or}\\& \targ & \qw\\\\}        &  \Qcircuit @C=1em @R=.7em {
& \gate{Y} & \qw\\}        & \Qcircuit @C=1em @R=.7em {
& \gate{Z} & \qw\\}         &   \Qcircuit @C=1em @R=.7em {
& \gate{H} & \qw\\}        & \Qcircuit @C=1em @R=.7em {
& \gate{S} & \qw\\}   & \Qcircuit @C=1em @R=.7em {
& \gate{T} & \qw\\}   &  \Qcircuit @C=1em @R=.7em {
& \ctrl{1} & \qw\\& \targ & \qw\\}     & \Qcircuit @C=1em @R=.7em {
& \ctrl{1} & \qw\\& \gate{Z} & \qw\\}&\Qcircuit @C=1em @R=.7em {& \ctrl{2} & \qw\\& \ctrl{1} & \qw\\& \targ & \qw\\} \\\hline
\end{tabular}
\end{table}

\subsection{Evaluation Metrics }

The performance of quantum design can be assessed using a variety of metrics such as Quantum Volume~\cite{cross2019validating}. Since this chapter emphasizes the evaluation of arithmetic circuits for quantum cryptanalysis, we only focus on three key metrics that are most relevant at the quantum circuit design level, as illustrated in Table~\ref{Metrics}.
\vspace{-0.5cm}
\begin{table}[ht]
\centering
\caption{Key Evaluation Metrics.\label{Metrics}}
\begin{tabular}{c|c|c}
\hline
{Metric} & {Significance}& {Description} \\
\hline
{Toffoli-Depth} &Time Complexity
& Number of Toffoli Layers
 \\\hline
{Toffoli-Count} & Gate Complexity 
& Total Number of Toffoli Gates
\\\hline
{Qubit-Count} & Space Complexity
& Total Qubits Required
 \\
\hline
\end{tabular}
\end{table}

\vspace{-1.3cm}
\subsection{Measurement-based Uncomputation (MBU)}

Uncomputation plays a critical role in quantum circuits, as it is required to clean up ancillas that hold intermediate values. 
There are two main approaches to uncomputation. The first is \textit{computation-based uncomputation}~\cite{Bennett1973LogicalRO}. This method involves copying the output state into ancilla register using CNOTs, followed by running the full inverse of the computing circuit. However, computation-based method is usually expensive, as it doubles the depth and gate count of the computing circuit.
The alternative way is \textit{measurement-based uncomputation (MBU)}~\cite{Gidney2018halvingcostof,gidney2019approximate,kornerup2021tight,luongo2025measurement}, which leverages intermediate measurements and classical feedback to probabilistically clean the ancillas. This general plug-and-play approach has gained traction in recent years, with the Logical-AND technique~\cite{Gidney2018halvingcostof} emerging as a widely used MBU method for Toffolis in various quantum arithmetic designs~\cite{PhysRevA.111.052611,wang2023efficient}.

\begin{figure}
\setlength{\abovecaptionskip}{0cm} 
\setlength{\belowcaptionskip}{-1cm} 
    \centering
    \includegraphics[width=0.8\linewidth]{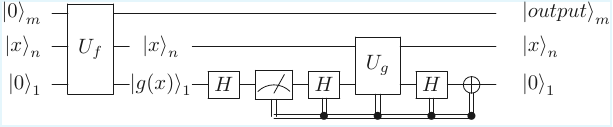}
    \caption{Example of Single-qubit Measurement-based
Uncomputation}
    \label{fig:MBU}
\end{figure}

As illustrated in Figure~\ref{fig:MBU}, we describe the single-qubit measurement-based uncomputation proposed by Luongo et al.~\cite{luongo2025measurement}.
This method naturally extends to multiple ancillas by sequentially applying the single-qubit method to each ancilla in turn.
Let $g:\{0,1\}^n \to \{0,1\}$ be a function, and let $U_g$ denote the self-adjoint unitary defined by $U_g \, |x\rangle_A |b\rangle_G = |x\rangle_A |b \oplus g(x)\rangle_G$,
for all $x \in \{0,1\}^n$ and $b \in \{0,1\}$. Suppose we start with a state in which the ancilla register $G$ contains the function value,
$\sum_x \alpha_x |x\rangle_A |g(x)\rangle_G$.

\medskip
\noindent\textbf{Step 1.} Apply the Hadamard gate to $G$, resulting in
    \begin{align}
    \sum_x \alpha_x |x\rangle_A \frac{|0\rangle + (-1)^{g(x)}|1\rangle}{\sqrt{2}}_G\label{formula: MBU_0}
    \end{align}

\medskip
\noindent\textbf{Step 2.} Measure $G$.
\begin{itemize}
    \item If the outcome is $0$ (probability 50\%), the state collapses to
    $\sum_x \alpha_x |x\rangle_A |0\rangle_G$,
    completing the uncomputation.
    \item If the outcome is $1$ (probability 50\%), the post-measurement state collapses to
    $\sum_x \alpha_x (-1)^{g(x)} |x\rangle_A |1\rangle_G$.
\end{itemize}

\medskip
\noindent\textbf{Step 3.} In the second case, apply a Hadamard gate on $G$ again.
    \begin{align}
    \sum_x \alpha_x (-1)^{g(x)} |x\rangle_A \frac{|0\rangle - |1\rangle}{\sqrt{2}}_G\label{formula: MBU_1}
    \end{align}

\medskip
\noindent\textbf{Step 4.} Apply $U_g$. The overall state becomes
    \begin{align}
    \sum_x \alpha_x (-1)^{g(x)} |x\rangle_A \frac{|g(x)\rangle - |1 \oplus g(x)\rangle}{\sqrt{2}}_G
=\sum_x \alpha_x |x\rangle_A |-\rangle_G\label{formula: MBU_2}
    \end{align}
\medskip
\noindent\textbf{Step 5.} Finally, apply a Hadamard gate followed by a NOT gate on $G$. The final state is
$\sum_x \alpha_x |x\rangle_A |0\rangle_G$,
which completes the uncomputation process.
\medskip

Computation-based uncomputation provides straightforward resource estimation but can be resource-intensive. In contrast, measurement-based uncomputation reduces circuit depth and gate counts but requires additional overhead from intermediate measurements and classical feedback processes. 
This trade-off has been explored in prior works~\cite{gidney2019windowed,luongo2025optimizing}, demonstrating that careful choice of uncomputation strategy depends on the specific resource constraints.

\vspace{-0.5cm}



\subsection{Conditionally Clean Ancilla}
In quantum circuit design, ancilla qubits can be used as temporary workspace to reduce the number of gates. Nevertheless, allocating clean ancilla is expensive, since each clean ancilla requires a dedicated qubit.
To address this problem, conditionally clean ancilla~\cite{nie2024quantum,khattar2024rise,PhysRevA.111.052611} has been introduced as a hybrid quantum resource bridging the gap between clean and dirty ancilla. Specifically, a conditionally clean ancilla is formally a dirty ancilla, which starts in an unknown state and is required to be restored to that state after use. Under specific conditions on other qubits, it is guaranteed to be in a known state. 

\begin{figure}[ht!]
\setlength{\abovecaptionskip}{0cm} 
\setlength{\belowcaptionskip}{-0.5cm} 
    \centering
    \subfigure[Clean Ancilla Design \label{fig:CCCCX_CA}]{
    \includegraphics[width=0.39\linewidth]{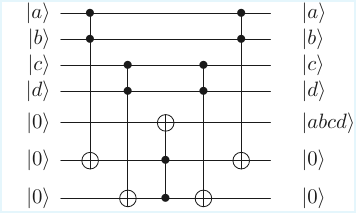}}
    \subfigure[Conditionally Clean Ancilla Design \label{fig:CCCCX_CCA}]{
    \includegraphics[width=0.55\linewidth]{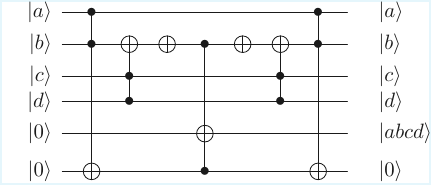}}
    \caption{Quantum $C^4$X Gate Implementations \label{fig:CCA_Example}}
\end{figure}

As an example, we consider the quantum design of a $C^4$X gate. Using only clean ancilla (Figure~\ref{fig:CCCCX_CA}) requires two ancillas, whereas employing conditionally clean ancilla in the decomposition (Figure~\ref{fig:CCCCX_CCA}) reduces the overall Qubit-Count. However, because conditionally clean ancilla must be restored to original states, measurement-based uncomputation cannot be applied to them.

\section{Computation-based Quantum Arithmetic}


In fault-tolerant quantum computing, the Clifford+T gate set serves as the standard universal basis for circuit design. Accordingly, this chapter focuses on the design of Clifford+T-based quantum arithmetic circuits. 
\subsection{Basic Arithmetic Operations}
\subsubsection{Addition\label{sec: add}}
Addition is one of the most fundamental operations in quantum computing. Similar to their classical counterparts, quantum adders can be categorized into several architectures, including ripple-carry, carry-lookahead~\cite{rosenberger1957simultaneous}, carry-save~\cite{earle1967latched}, and hybrid designs, as summarized in Table~\ref{table: c+T_Add}.
\vspace{-0.5cm}
\begin{table}[!ht]
\caption{Overview of Prominent Quantum Adders.}
\label{table: c+T_Add}
\setlength{\abovecaptionskip}{-1cm} 
\setlength{\belowcaptionskip}{-1cm} 
\centering
\scalebox{0.9}{
\begin{tabular}{l|l|l|l}
\hline
Year &Article &Category & Key Innovations \\
\hline
1996 & Vedral et al.~\cite{VBE}&\textcolor{color_ripple_carry}{Ripple-Carry}&
First Quantum Ripple-Carry Adder 
\\ \hline
1998&Gossett~\cite{gossett1998quantum}&\textcolor{orange}{Carry-Save}& First Quantum Carry-Save Adder\\ \hline
2004 & Cuccaro et al.~\cite{Cuccaro}&\textcolor{color_ripple_carry}{Ripple-Carry} & Use Only One Ancilla\\ \hline
2004 & Draper et al.~\cite{Draper}&\textcolor{color_carry_look_ahead}{Carry-Lookahead}& First Quantum Carry-Lookahead Adder\\ \hline
2005 & Takahshi et al.~\cite{Takahashi05}& \textcolor{color_ripple_carry}{Ripple-Carry} &Use Zero Ancilla\\ \hline
2008 & Takahshi et al.~\cite{Takahashi08}& \textcolor{color_hybrid}{Hybrid}&Design Ripple-Carry \& Carry-Lookahead Adder\\ \hline
2009 & Takahshi et al.~\cite{Takahashi09}& \textcolor{color_ripple_carry}{Ripple-Carry}& Reduce  Size \& Depth of Adder without Ancilla \\ \hline
2009 & Takahshi et al.~\cite{Takahashi09}& \textcolor{color_hybrid}{Hybrid}& Improved Design over~\cite{Takahashi08}\\ \hline
2009 & Takahshi et al.~\cite{Takahashi09}& \textcolor{color_hybrid}{Modified Hybrid}& Improved Design over~\cite{Takahashi09} \\ \hline
2016 & Wang et al.~\cite{paper4_2016} & \textcolor{color_ripple_carry}{Ripple-Carry} & Simplify  Building Blocks of Ripple-Carry Adder \\ \hline
2018 & Gidney~\cite{Gidney2018halvingcostof}&\textcolor{color_ripple_carry}{Ripple-Carry}& Introduce Logical-AND to Reduce  Cost\\ \hline
2021 &Gayathri et al.~\cite{paper3_2021}&\textcolor{color_ripple_carry}{Ripple-Carry}&Simplify  Building Blocks of Ripple-Carry Adder\\ \hline
2023 &Wang et al.~\cite{wang2023higher} & \textcolor{color_hybrid}{Hybrid}&Simplify  Building Blocks of Hybrid Adder \\ \hline
2023 &Wang et al.~\cite{wang2023reducing} & \textcolor{color_carry_look_ahead}{Carry-Lookahead}& Propose Quantum Ling Base Adder\\ \hline
2025 &Wang et al.~\cite{wang2025optimal} &\textcolor{color_carry_look_ahead}{Carry-Lookahead} &Explore Parallel Prefix Trees in Quantum \\ \hline
2025 &Gu et al.~\cite{gu2025resource} &\textcolor{color_carry_look_ahead}{Carry-Lookahead}  &Hybrid Architecture based on~\cite{Draper} and~\cite{wang2025optimal}\\ \hline
2025 &Remaud et al.~\cite{remaud2025ancilla} &\textcolor{color_ripple_carry}{Ripple-Carry} &Achieve Sublinear Depth without Ancilla\\ \hline
2025 &Kim et al.~\cite{kim2025tree}&\textcolor{orange}{Carry-Save}&Improved Design over~\cite{gossett1998quantum}\\
\hline
\end{tabular}}
\end{table}
\vspace{-0.5cm}
\begin{enumerate}
    \item \textbf{Ripple-Carry Structure.} 
    The ripple-carry structure represents the earliest class of quantum adders. In 1996, Vedral et al.~\cite{VBE} introduced explicit quantum circuits capable of performing fundamental arithmetic operations, including the first $n$-bit ripple-carry adder, marking a significant milestone in quantum arithmetic.
    Subsequent refinements focused on reducing ancillary qubit usage and circuit complexity. In 2004, Cuccaro et al.~\cite{Cuccaro} proposed a linear-depth ripple-carry adder requiring only a single ancillary qubit, achieving reduced depth and gate count. The following year, Takahashi et al.~\cite{Takahashi05} developed the first ancilla-free ripple-carry adder, and in 2009~\cite{Takahashi09}, they further optimized its size to $7n-6$, the smallest known for addition without ancillas at that time.
    Further improvements focused on gate efficiency. In 2016, Wang et al.~\cite{paper4_2016} reduced the number of Toffoli gates per output bit from at least two to one, while Gidney~\cite{Gidney2018halvingcostof} employed measurement-based uncomputation, halving the overall cost of addition by reducing the T-gate usage in Toffoli pairs. Moreover, Gayathri et al.~\cite{paper3_2021} enhanced the full-adder structure and applied a reversible pebble-game-based uncomputing scheme~\cite{Bennett1973LogicalRO} to minimize redundant Toffoli gates.
    Most recently, Remaud et al.~\cite{remaud2025ancilla} introduced a sublinear-depth ripple-carry adder without ancilla qubits, utilizing conditionally cleaned ancillas, representing the current state of the art in ripple-carry quantum addition.

    \item \textbf{Carry-Lookahead Structure.} The first quantum carry-lookahead adder was proposed by Draper et al.~\cite{Draper}, which reduced the Toffoli-Depth from O(n) for ripple-carry adders to $O(\log n)$, with only a modest increase in qubit requirements.
    In 2023, Wang et al.~\cite{wang2023reducing} introduced the Ling basis in quantum carry-lookahead addition. While the initial quantum Ling adder offered only limited improvement, it highlighted the potential to modify the propagation and generation base within carry-lookahead designs.
    Subsequently, Wang et al.~\cite{wang2025optimal} systematically explored 160 alternative carry-propagation structures to determine the optimal depth configuration. By implementing the Sklansky prefix tree~\cite{Sklansky}, they achieved a Toffoli-Depth of $\log  n+O(1)$, effectively halving the depth of Draper's design~\cite{Draper}. Therefore, their design is also commonly referred to as the quantum optimal-depth adder.
    Building upon the optimal-depth adder, Gu et al.~\cite{gu2025resource} proposed a carry-lookahead adder that achieves $\log n +\log \log n +O(1)$ Toffoli-Depth  with only $O(n)$ ancillas, providing a balanced trade-off between depth and Qubit-Count. Nevertheless, the lowest achievable Toffoli-Depth remains $\log n+O(1)$, as realized by the optimal-depth adder~\cite{wang2025optimal}.

    \item \textbf{Carry-Save Structure.}
    The first quantum carry-save structure was introduced by Gossett in 1998~\cite{gossett1998quantum}. While foundational, this early work leaves significant opportunities for optimization. 
    Building on this initial work, Kim et al.~\cite{kim2025tree} proposed a tree-based quantum carry-save adder, which further integrates Wallace and Dadda trees to improve carry propagation in multi-operand additions.
    Although carry-save structures have received less attention than other addition structures so far, they are widely used in modular multiplication and modular exponentiation circuits, as detailed in~\cite{wang2024minimum,wang2024quantumbook}.

    \item \textbf{Hybrid Structure.} As discussed previously, ripple-carry adders require fewer qubits but exhibit a large Toffoli-Depth~\cite{VBE,Cuccaro}, whereas carry-lookahead adders reduce Toffoli-Depth at the expense of increased qubit requirements~\cite{Draper}. 
    To combine the advantages of both structures, the first hybrid adder was proposed by Takahashi et al.~\cite{Takahashi08} in 2008, integrating a modified carry-lookahead adder based on~\cite{Draper} with parallel applications of a ripple-carry adder~\cite{Takahashi05}. 
    This design was subsequently generalized and simplified in 2009~\cite{Takahashi09}, achieving logarithmic-depth circuits with fewer qubits and enhanced efficiency through the use of unbounded fan-out gates.
    More recently, Wang et al.~\cite{wang2023higher} proposed a higher-radix carry-lookahead adder, which separates the propagation and summation phases within the Manchester Carry Chain. This design enables flexible carry and sum paths, providing improved efficiency and adaptability in various scenarios.

\item \textbf{Classical–Quantum Structure.}
Beyond the four addition structures discussed above, several studies~\cite{Takahashi09,haner2016factoring,gidney2025classical} have proposed simplified approaches that treat the operation as addition by a constant, also referred to as quantum constant addition or classical–quantum addition. Compared to general quantum adders, these designs require fewer qubits but rely more heavily on classically controlled quantum operations.
\end{enumerate}

\subsubsection{Subtraction\label{sec: sub}}
Compared to quantum adders, quantum subtraction has received less attention. However, as a fundamental arithmetic operation, it remains essential. The construction of quantum subtractors can be categorized into three primary approaches, as detailed below.
\vspace{-0.5cm}
\begin{figure}[ht!]
\setlength{\abovecaptionskip}{0cm} 
\setlength{\belowcaptionskip}{-1cm} 
    \centering
    \subfigure[Ripple-Borrow Subtraction\label{Approach1_Sub}]{\includegraphics[width=0.9\linewidth]{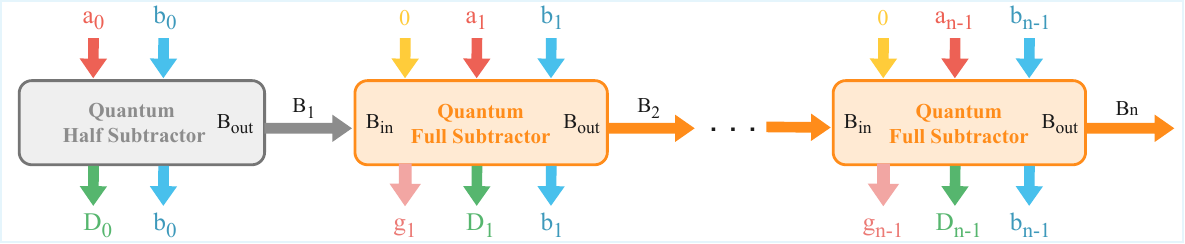} 
    }
    \subfigure[Without Input Carry\label{Approach2_Sub}]{\includegraphics[width=0.43\linewidth]{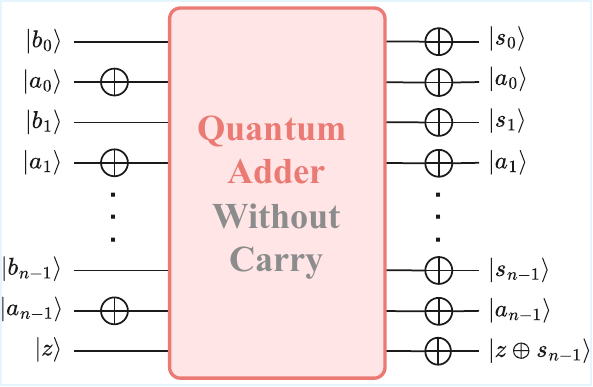}}
    \hfill
    \subfigure[With Input Carry\label{Approach3_Sub}]{\includegraphics[width=0.41\linewidth]{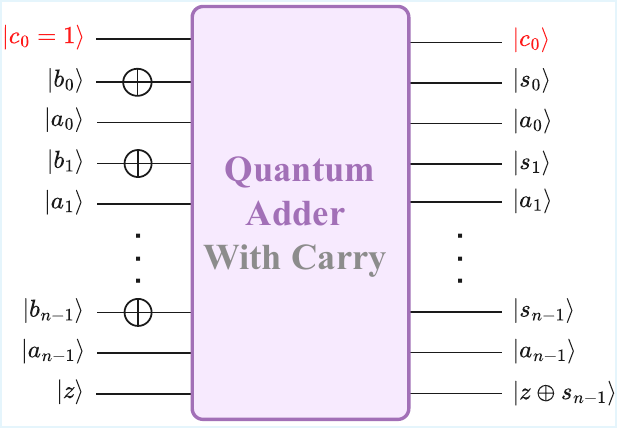}}
    \hfill
    \caption{Structures of $n$-bit Quantum Subtractor.\textcolor{blue}{\textbf{Anubhab:} We can redraw these diagrams in tikz}}
\end{figure}

\begin{enumerate}
    \item \textbf{Ripple-Borrow Subtraction.}
    Building upon the quantum half-subtractor and full-subtractor, the ripple-borrow approach offers a straightforward method for constructing an $n$-bit quantum subtractor. This approach has been widely adopted in existing designs~\cite{cheng2002quantum,thapliyal2009design,thapliyal2016mapping}. As illustrated in Figure~\ref{Approach1_Sub}, the ripple-borrow subtractor shares a similar structure with quantum ripple-carry adders.
    \item \textbf{Subtraction Without Input Carry.} Apart from constructing quantum subtractors directly from half and full subtractor building bolcks, an $n$-bit quantum subtractor can also be realized by introducing minor modifications to an existing $n$-bit quantum adder circuit. According to Formula~\ref{formula: sub0}, this can be efficiently achieved through the insertion of a series of quantum NOT gates. Specifically, the input register $a$ is complemented at the beginning of the computation, while both $a$ and the final sum are complemented at the end. As illustrated in Figure~\ref{Approach2_Sub}, the overall structure of this subtractor is closely similar to the underlying quantum adder.
    \begin{align}
    &A-B=(A'+ B)'\label{formula: sub0}
    \end{align}
    \item \textbf{Subtraction With Input Carry.} 
    Another formulation~\ref{formula: sub1} also can be used to modify a quantum adder to implement an $n$-bit quantum subtractor. This approach employs an $n$-bit quantum adder with an input carry, as illustrated in Figure~\ref{Approach3_Sub}. Specifically, the input register $b$ is complemented at the beginning, and the input carry $c_0$ is set to 1. This method is similar to Method 2, but differs slightly in the number of NOT gates and qubit usage—requiring one additional qubit for the carry-in while using $(2n+1)$ fewer NOT gates than Method 2. Some studies treat this carry-in as an increment operation~\cite{luongo2025measurement} and implement it using a separate increment circuit~\cite{Gidney_Increment}. However, implementing it directly as a carry-in avoids the additional overhead associated with a separate increment circuit.
    \begin{align}
    &A-B=A+ B'+1\label{formula: sub1}
    \end{align}
\end{enumerate}


\subsubsection{Multiplication}
Numerous efficient classical multiplication algorithms have been developed, including the Shift-and-Add, Karatsuba~\cite{karatsuba1962multiplication}, Wallace Tree~\cite{wallace1964suggestion}, and Toom–Cook~\cite{cook1969minimum} methods. These classical algorithms form the basis for many quantum counterparts, as their core principles of decomposition and partial product accumulation remain applicable in the quantum domain. Consequently, quantum multiplication circuits can also be classified according to these classical categories, as summarized in Table~\ref{tabCliffordPlusTMultiplier}.

\begin{table}[!ht]
\caption{Overview of Prominent Quantum Multipliers.}
\label{tabCliffordPlusTMultiplier}
\setlength{\abovecaptionskip}{-1cm} 
\setlength{\belowcaptionskip}{-1cm} 
\centering
\resizebox{\textwidth}{!}{
\begin{tabular}{l|l|l|l}
\hline
Year &Article &Category & Key Innovations \\
\hline
1998&Zalka~\cite{zalka1998fast}&\textcolor{color_schonhage_strassen}{Sch{\"o}nhage-Strassen}& First Quantum Sch{\"o}nhage-Strassen Multiplier\\ \hline
2004&Parent et al.~\cite{parent2017improved}&\textcolor{color_shift_and_add}{Shift-and-Add}& First Shift-and-Add Multiplier\\ \hline
2004&Draper et al.~\cite{Draper}&\textcolor{color_shift_and_add}{Shift-and-Add}&Use $n$ Carry-Lookahead Adders\\ \hline
2014& Lin et al.~\cite{lin2014qlib}&\textcolor{color_shift_and_add}{Shift-and-Add}&Construct a Quantum Module Library\\ \hline
2015 &Kepley et al.~\cite{kepley2015quantum}&\textcolor{color_karatsuba}{Karatsuba}& First Quantum Karatsuba Multiplier\\ \hline
2016 & Jayashree et al.~\cite{jayashree2016ancilla}&\textcolor{color_shift_and_add}{Shift-and-Add}& Optimize Ancillas and Eliminate Garbage\\ \hline
2017&Parent et al.~\cite{parent2017improved}&\textcolor{color_karatsuba}{Karatsuba}&Optimize Qubit-Count for Karatsuba Multiplier\\ \hline
 2018&Dutta et al.~\cite{Toom-Cook-multi}&\textcolor{color_toom_cook}{Toom-Cook 2.5}& First Quantum Toom-Cook Multiplier\\ \hline
 2018&Mu{\~n}oz-Coreas et al.~\cite{munoz2018quantum}&\textcolor{color_shift_and_add}{Shift-and-Add}& Optimize Toffoli-Count and Toffoli-Depth\\ \hline
 2019&Gidney~\cite{gidney2019asymptotically}&\textcolor{color_karatsuba}{Karatsuba}& Use Inline-Mutation Calls to Optimize Qubit-Count\\ \hline
 2020&Van Hoof~\cite{van2019space}&\textcolor{color_karatsuba}{Karatsuba}& Optimize Qubit-Count for Karatsuba Multiplier\\ \hline
 2021&Larasati et al.~\cite{larasati2021quantum}&\textcolor{color_toom_cook}{Toom-Cook 3}&Optimize Qubit-Count and Toffoli-Depth\\ \hline
 2021& Gayathri et al.~\cite{gayathri2021t}&\textcolor{color_wallace_tree}{Wallace Tree}& First Quantum Wallace Tree Multiplier\\ \hline
 2022 & Sajadimanesh et al.~\cite{sajadimanesh2022practical}&\textcolor{color_shift_and_add}{Shift-and-Add}&Adapt  Design for Quantum Devices \\ \hline
 2023&  Putranto et al.~\cite{putranto2023space}& \textcolor{color_toom_cook}{Toom-Cook 8}&Optimize Space \& Time Cost\\ \hline
 2023& Orts et al.~\cite{orts2023improving}&\textcolor{color_wallace_tree}{Wallace Tree}&Support Arbitrary Size Multiplication\\ \hline
 2023&Nie et al.~\cite{nie2023quantum}&\textcolor{color_schonhage_strassen}{Sch{\"o}nhage-Strassen}&Improve Sch{\"o}nhage-Strassen Multiplier\\ \hline
 2024&  Wardhani et al.~\cite{wardhani2024high}& \textcolor{color_toom_cook}{Toom-Cook 8.5, 10.5, 20.5 }&High- and Half Degree Toom-Cook Design\\ \hline
 2025& Hwang et al.~\cite{10.1145/3672608.3707921}&\textcolor{color_shift_and_add}{Shift-and-Add}& A Configurable Approximate Multiplier\\ \hline
 2025& Wang et al.~\cite{wang2025reducing}&\textcolor{color_wallace_tree}{Wallace Tree}&Use Compressor as Primitive\\ \hline
\end{tabular}}
\vspace*{-10pt}
\end{table}

\begin{enumerate}
    \item \textbf{Shift-and-Add Structure.}
    This structure, also known as schoolbook multiplication, represents the most straightforward approach to performing multiplication, where the operation is decomposed into a sequence of additions to produce the final product. 
    Early quantum multipliers, such as those proposed by Draper et al.~\cite{Draper} and Parent et al.~\cite{parent2017improved}, adopted this technique. Despite its conceptual simplicity, the Shift-and-Add multiplication method remains inefficient due to its limited potential for parallelization.
    Lin et al.~\cite{lin2014qlib} introduced the QLib quantum module library, which provides various arithmetic modules for benchmarking quantum logic and synthesis. However, the Shift-and-Add multiplier module in QLib still requires a relatively high quantum cost.
    In 2016, Jayashree et al.~\cite{jayashree2016ancilla} refined the previous Shift-and-Add multiplier by applying reversible logic to create a garbage-free, ancilla-efficient design. This method reduced ancillary qubit requirements from $3N+1$ in Lin et al.’s design~\cite{lin2014qlib} to $2N+1$ for an $N$-bit multiplication, addressing one of the major inefficiencies in quantum arithmetic.
    Moreover, Mu{~n}oz-Coreas et al.~\cite{munoz2018quantum} proposed a T-count-optimized Shift-and-Add multiplier in 2018. By employing an improved conditional adder, their design achieved lower Toffoli-Count and Toffoli-Depth while maintaining the same ancilla qubits as Jayashree et al.~\cite{jayashree2016ancilla}'s design.
    In order to enable practical deployment, Sajadimanesh et al.~\cite{sajadimanesh2022practical} extended the Shift-and-Add framework to real NISQ hardware by employing approximate computing.
    Following this direction, Hwang et al.~\cite{10.1145/3672608.3707921} proposed a configurable approximate quantum Shift-and-Add multiplier that integrates approximate and exact controlled adders with a truncation scheme.  By accepting minor accuracy losses, these approximate multipliers significantly reduce resource costs while remaining effective on noisy quantum devices.
    \item \textbf{Karatsuba Structure.}
    This is a classical divide-and-conquer technique~\cite{karatsuba1962multiplication} that accelerates multiplication by recursively decomposing large operands into smaller subproblems. The inherent efficiency and recursive structure make it particularly well suited for adaptation in quantum computing.
    In 2015, Kepley et al.~\cite{kepley2015quantum} were the first to apply the Karatsuba multiplication algorithm in the quantum domain, achieving a T-gate complexity of $O(n^{\log 3})$. Their design reduced the Toffoli and Clifford gate counts compared to early quantum Shift-and-Add multipliers~\cite{parent2017improved,Draper}, though at the cost of increased Qubit-Count.
    By applying pebble-game optimization on ternary trees, Parent et al.~\cite{parent2017improved} optimized Kepley et al.~\cite{kepley2015quantum}'s work, thereby reducing the overall Qubit-Count without increasing Toffoli-Count or Toffoli-Depth. 
    In 2019, Gidney~\cite{gidney2019asymptotically} further improved the quantum Karatsuba multiplication by introducing recursive output reuse, significantly reducing space complexity to 
     O(n) while maintaining similar gate efficiency.
    Most recently, Van Hoof~\cite{van2019space} proposed a space-efficient variant of the quantum Karatsuba multiplier that requires only $3n$ qubits, achieving comparable Toffoli-count and depth performance to that of Kepley et al.~\cite{kepley2015quantum}, albeit with a moderate increase in CNOT-Count.
    
    \item \textbf{Toom-Cook Structure.}
    Building upon Karatsuba multiplication, this structure achieves improved performance by employing an extended divide-and-conquer strategy.
    \vspace{-0.5cm}
    \begin{table}[ht!]
    \caption{Asymptotic Performance Analysis of Quantum Toom-Cook Multipliers.}
\label{Toom-Cook-Compare}
\renewcommand{\arraystretch}{1.15}
\setlength{\abovecaptionskip}{-1cm} 
\setlength{\belowcaptionskip}{-1cm} 
    \centering
\begin{tabular}{l|l|l|l|l}
\hline
Article              & Structure     & Qubit-Count          & Toffoli-Count        & Toffoli-Depth        \\ \hline
 Putranto et al.~\cite{putranto2023space} & \textcolor{color_toom_cook_2}{Toom-Cook 2}   & $O(n^{1.589})$ & $O(n^{\log_2{3}})$  & $O(n^{1.217})$\\ \hline
 Dutta et al.~\cite{Toom-Cook-multi}& \textcolor{color_toom_cook_2}{Toom-Cook 2.5} &    $O(n^{1.404})$& $O(n^{\log_6 {16}})$ &$O(n^{1.143})$\\ \hline
Larasati et al.~\cite{larasati2021quantum} & \textcolor{color_toom_cook_2_5}{Toom-Cook 3}   & $O(n^{1.35})$ & $O(n^{2})$ &$O(n^{1.112})$\\ \hline
Putranto et al.~\cite{putranto2023space}& \textcolor{color_toom_cook_3}{Toom-Cook 4}  &$O(n^{1.313})$&$O(n^{\log_4 {7}})$&$O(n^{1.09})$\\ \hline
Putranto et al.~\cite{putranto2023space}& \textcolor{color_toom_cook_4}{Toom-Cook 8}   &$O(n^{1.245})$ &$O(n^{\log_8 {15}})$ &$O(n^{1.0569})$\\ \hline
Wardhani et al.~\cite{wardhani2024high}& \textcolor{color_toom_cook_4}{Toom-Cook 8.5} &$O(n^{1.236})$ &$O(n^{\log_9 {17}})$ &$O(n^{1.053})$\\ \hline
Wardhani et al.~\cite{wardhani2024high}& \textcolor{color_toom_cook_8}{Toom-Cook 10.5} &$O(n^{1.222})$ &$O(n^{\log_{11} {21}})$ &$O(n^{1.047})$\\ \hline
Wardhani et al.~\cite{wardhani2024high}& \textcolor{color_toom_cook_8}{Toom-Cook 20.5} &$O(n^{1.186})$ &$O(n^{\log_{21} {41}})$ &$O(n^{1.033})$\\ \hline
\end{tabular}
\end{table}
    In 2018, Dutta et al.~\cite{Toom-Cook-multi} proposed the first quantum Toom–Cook multiplier, based on Toom–2.5. By employing reversible pebble games to uncompute intermediate results, their design achieved improvements in Toffoli-Depth, Toffoli-Count, and Qubit-Count. 
    Subsequently, Larasati et al.~\cite{larasati2021quantum} extended this design to Toom–3 multiplication, achieving lower asymptotic Toffoli-Depth and Qubit-Count compared to the quantum Toom–2.5 design~\cite{Toom-Cook-multi}, though at the cost of a higher Toffoli gate count due to the division operation.
    Further developments have focused on higher-degree Toom–Cook multipliers to further optimize quantum resource usage. Specifically, Putranto et al.~\cite{putranto2023space} introduced a Toom–8-way quantum multiplier, minimizing both asymptotic cost and implementation complexity. In 2024, Wardhani et al.~\cite{wardhani2024high} investigated Toom–8.5-way, 10.5-way, and 20.5-way multipliers, targeting high- and half-degree quantum multiplication.
    Among all of these quantum Toom-Cook Multipliers, the Toom–20.5 multiplier demonstrated the highest efficiency, achieving significant reductions in Qubit-Counts, Toffoli-Counts, and Toffoli-Depth compared to previous Toom–Cook-based designs, as summarized in Table~\ref{Toom-Cook-Compare}. 
    \item \textbf{Sch{\"o}nhage-Strassen Structure.}
    In classical computing, the Sch{\"o}nhage-Strassen structure outperforms both the Karatsuba and Toom-Cook structures, motivating its exploration in quantum computing. 
    In 1998, the initial quantum Sch{\"o}nhage-Strassen multiplier was proposed by Zalka~\cite{zalka1998fast}, utilizing a second-level FFT to reduce the quantum cost of integer multiplication.
    More recently, Nie et al.~\cite{nie2023quantum} developed a family of quantum Sch{\"o}nhage-Strassen multiplication circuits. Compared with existing quantum Toom-Cook~\cite{Toom-Cook-multi} and Karatsuba~\cite{gidney2019asymptotically} multipliers, their design reduces both gate count and circuit depth, but with higher qubit requirements.
    \item \textbf{Wallace Tree Structure.}
    As one of the fastest and most widely adopted multiplication architectures, the Wallace tree structure~\cite{wallace1964suggestion} achieves high computation efficiency by hierarchically reducing partial products through cascaded layers of full and half adders, thereby minimizing the overall time cost.  
    Based on this structure, Gayathri et al.~\cite{gayathri2021t} introduced the first quantum Wallace tree multiplier in 2021. 
    Subsequently, Orts et al.~\cite{orts2023improving} extended this approach by developing a scalable quantum Wallace tree circuit capable of handling arbitrary operand sizes. Their design refined low-level gate operations and optimized resource utilization, achieving superior performance compared to the earlier work by Gayathri et al.~\cite{gayathri2021t}.
    In 2025, Wang et al.~\cite{wang2025reducing} identified that most quantum Wallace tree multipliers rely on basic half- and full-adder components, which limit the overall efficiency. Therefore, they introduced a generalized $(m:k)$ compressor-based Wallace tree architecture. By applying brute-force and dynamic programming optimization, their design achieved more than 50\% reduction in T-depth and T-count, while maintaining a comparable Qubit-Count. This work demonstrates the potential of compressor primitives to significantly enhance the performance of quantum arithmetic designs.
\end{enumerate}

\subsubsection{Division} 

In classical computing, division algorithms are generally classified into fast and slow categories. Since most quantum division algorithms are adapted from their classical counterparts, a similar classification can be established in the quantum domain, as summarized in Table~\ref{tabCliffordPlusTDivision}.
\vspace{-0.5cm}
\begin{table}[!h]
\caption{Overview of Prominent Quantum Dividers.}
\label{tabCliffordPlusTDivision}
\setlength{\abovecaptionskip}{-1cm} 
\setlength{\belowcaptionskip}{-1cm} 
\centering
\scalebox{0.9}{
\begin{tabular}{l|l|l|l}
\hline
Year &Article &Category & Key Innovations \\
\hline
2019 & Thapliyal et al.~\cite{thapliyal2019quantum}& \textcolor{color_restoring}{Restoring}&The First Quantum Restoring Divider\\ \hline
2019 & Thapliyal et al.~\cite{thapliyal2019quantum}& \textcolor{color_non_restoring}{Non-Restoring}&The First Quantum Non-Restoring Divider\\ \hline
2021 & Gayathri et al.~\cite{gayathri2021t}&\textcolor{color_goldschmidt}{Goldschmidt} &The First Quantum Goldschmidt Divider\\ \hline
2022 & Gayathri et al.~\cite{gayathri2022efficient}&\textcolor{color_newton_raphson}{Newton-Raphson} &The First Quantum Newton-Raphson Divider\\ \hline
2022 & Li et al.~\cite{li2022circuit} &\textcolor{color_long_division}{Long Division} & Based on Approximate Gate Set\\ \hline
2022 & Yuan et al.~\cite{yuan2022novel}&\textcolor{color_long_division}{Long Division}&Reduce Overall Cost of Designs~\cite{thapliyal2019quantum}\\ \hline
2024& Wang et al.~\cite{wang2024boosting} &\textcolor{color_non_restoring}{Non-Restoring}&Apply Comprehensive DSE to Reduce Cost\\ \hline
2024& Sajadimanesh et al.~\cite{PhysRevA.109.052601}&\textcolor{color_restoring}{Restoring}&Implement on Quantum Computers\\ \hline
2024& Sajadimanesh et al.~\cite{PhysRevA.109.052601}&\textcolor{color_non_restoring}{Non-Restoring}&Implement on Quantum Computers\\ \hline
2024& Orts et al.~\cite{orts2024quantum}&\textcolor{color_long_division}{Long Division}&Reduce Toffoli-Count and Toffoli-Depth\\
\hline
\end{tabular}}
\vspace*{-4pt}
\end{table}

\begin{enumerate}
    \item \textbf{Slow Division.} As the name suggests, slow division algorithms emphasize conceptual simplicity rather than computational speed. Classical representatives of this category include long division, restoring division, non-restoring division~\cite{shaw1950arithmetic}, and SRT division~\cite{harris1998srt}. Although these algorithms are less efficient compared to fast division algorithms, their straightforward architectures make them highly suitable for investigating quantum implementations of division.
    The first quantum realizations of slow division were proposed by Thapliyal et al.~\cite{thapliyal2019quantum}, who developed quantum restoring and non-restoring dividers inspired by their classical counterparts. However, as the T-depth of their design was evaluated using a non-standard approach, its reported cost is not directly comparable with other designs. A corrected analysis was later provided in~\cite{wang2024boosting}.
    Quantum implementations of long division were first investigated in 2022, when Li et al.~\cite{li2022circuit} introduced the use of approximate quantum gates optimized for reduced T-depth and T-count. Leveraging these gates, they constructed efficient arithmetic modules—including a quantum divider based on the long division algorithm. 
    In the same year, Yuan et al.~\cite{yuan2022novel} proposed a fault-tolerant quantum divider based on quantum comparators and subtractors. Both works achieved significant reductions in Toffoli-count, Toffoli-depth, and Qubit-Count compared to existing quantum restoring and non-restoring dividers~\cite{thapliyal2019quantum}.
    Based on previous quantum slow dividers, further refinements were made by Wang et al.~\cite{wang2024boosting}, who focused on design space exploration (DSE) of slow division architectures. By optimizing key arithmetic sub-blocks and systematically evaluating trade-offs across different design parameters, their work achieved measurable improvements in both depth and resource efficiency.
    Moreover, Sajadimanesh et al.~\cite{PhysRevA.109.052601} demonstrated practical implementations of quantum restoring and non-restoring dividers on NISQ hardware. Their approach combined dynamic circuits with approximate computing, leveraging mid-circuit measurement to reduce qubit usage and mitigate hardware noise. These circuits were successfully executed on IBM quantum devices, providing one of the earliest hardware demonstrations of quantum division.
    Most recently, Orts et al.~\cite{orts2024quantum} proposed an optimized quantum long division circuit that minimizes T-gate count while maintaining a comparable qubit cost. By leveraging T-optimized circuit variants, their design achieves superior Toffoli-Depth and Toffoli-Count performance compared with previous quantum slow division implementations~\cite{thapliyal2019quantum,yuan2022novel}.
    \item \textbf{Fast Division.} In contrast to slow division, fast division aims to minimize computational latency in calculating the quotient and remainder. Within this category, the most representative approaches are the Newton–Raphson and Goldschmidt methods~\cite{goldschmidt1964applications}, both of which employ iterative techniques to achieve rapid convergence toward the final result.
    Building on these classical foundations, Gayathri et al.~\cite{gayathri2021t} proposed the first quantum fast division circuit based on the Goldschmidt algorithm in 2021, adopting the IEEE 754 single-precision floating-point representation. Despite its novelty, this implementation incurred a relatively high overall quantum cost.
    In a subsequent work, Gayathri et al.~\cite{gayathri2022efficient} introduced an improved quantum divider based on the Newton–Raphson method. Compared to existing fast division circuits that rely on the Goldschmidt algorithm~\cite{gayathri2021t,ananthalakshmi2017novel}, their design achieved significant reductions in both T-count and Qubit-Count. Overall, among the existing quantum fast division architectures, the Quantum Newton–Raphson Divider currently represents the most efficient structure.
\end{enumerate}

\subsection{Composite Operations}
After introducing the basic quantum arithmetic components, we then provide an overview of the key composite arithmetic operations used in Shor’s cryptanalysis of RSA and ECC, as summarized in Table~\ref{tableModExp}.

\begin{table}[!h]
\caption{Overview of Prominent Quantum Composite Arithmetic Designs.}
\label{tableModExp}
\setlength{\abovecaptionskip}{-1cm} 
\setlength{\belowcaptionskip}{-1cm} 
\centering
\resizebox{\textwidth}{!}{
\begin{tabular}{l|l|l|l}
\hline
Year &Article &Composite Operation & Key Innovations \\
\hline
1996&Vedral et al.~\cite{VBE}&\textcolor{color_Mod_Exp}{Modular Exponentiation}&Use Arithmetic for Modular Exponentiation\\ \hline
1996&Beckman et al.~\cite{beckman1996efficient}&\textcolor{color_Mod_Exp}{Modular Exponentiation}&Use Arithmetic for Modular Exponentiation\\ \hline
2005&Van Meter et al.~\cite{van2005fast}&\textcolor{color_Mod_Exp}{Modular Exponentiation}&Boost the Efficiency with Concurrency\\ \hline
2012&Markov et al.~\cite{markov2012constant}&\textcolor{color_Mod_Exp}{Modular Exponentiation}&Use Predefined Parameters for Specific Cases\\ \hline
2012 & Amento et al. \cite{amento2012efficient} & \textcolor{color_point_add}{Point Addition}&Use Inversion-Free Projective Point Addition\\ \hline
2017& Roetteler et al. \cite{roetteler2017quantum}&\textcolor{color_point_add}{Point Addition} & Use Modular Arithmetic for Point Addition \\ \hline
2018&H{\"a}ner et al.~\cite{haner2018optimizing}&\textcolor{color_Mod_Exp}{Modular Exponentiation}&Apply Piecewise Polynomial Approximation\\ \hline
2019& Das et al.~\cite{das2019optimizing}&\textcolor{color_Mod_Exp}{Modular Exponentiation}&Map Logic Designs to Quantum Architectures\\ \hline
2020& Banegas et al. \cite{banegas2020concrete}&\textcolor{color_point_add}{Point Addition}& Use In-Place Point Addition Structure\\ \hline
2021& Häner et al. \cite{haner2020improved}&\textcolor{color_point_add}{Point Addition}& Apply Window Technique \\ \hline
2025& Jang et al. \cite{jang2025new}&\textcolor{color_point_add}{Point Addition}& Propose Out-of-Place and Improved In-Place Design  \\ \hline

\end{tabular}}
\vspace*{-4pt}
\end{table}

\subsubsection{Quantum Modular Exponentiation\label{sec: mod}}
In Shor’s algorithm for breaking RSA, modular exponentiation dominates the overall computational cost. This composite operation is constructed from quantum modular multiplication and modular addition~\cite{jang2023quantum,kim2024toffoli,zhang2025optimized,wang2024minimum,luongo2025measurement}. For conciseness, here we concentrate on modular exponentiation itself, rather than the basic arithmetic blocks on which it depends.

Early research in quantum modular exponentiation began with computation-based designs. In 1996, Vedral et al.~\cite{VBE} presented the first explicit quantum networks for arithmetic, constructing modular exponentiation from quantum ripple-carry adders. Their approach offered conceptual clarity but inherited the slow performance of ripple-carry addition and required ancilla qubits that scaled linearly with input size. 
Based on this initial design, Beckman et al.~\cite{beckman1996efficient} developed a similar modular exponentiation architecture, incorporating more classical computation and exploring various time–space trade-offs to optimize the overall structure.

Subsequent work shifted toward improving execution time. 
Specifically, Van Meter et al.~\cite{van2005fast} introduced logical concurrency to parallelize arithmetic operations in modular exponentiation, showing that improved adders and concurrent gate execution can substantially accelerate general modular exponentiation. 
Another research examined specialized instances of the problem. In 2012, Markov et al.~\cite{markov2012constant} constructed optimized modular exponentiation circuits with specific input cases, achieving constant-factor improvements in quantum cost.
In 2018, a different approach was explored by H{\"a}ner et al.~\cite{haner2018optimizing}, who implemented modular exponentiation using piece-wise polynomial approximations. Their method employed basic computation-based arithmetic blocks to evaluate many polynomials in parallel, resulting in an efficient modular exponentiation framework. 
Complementary to these analytical constructions, Das et al.~\cite{das2019optimizing} demonstrated that practical logic-synthesis workflows can generate competitive quantum modular exponentiation implementations by translating Verilog designs into reversible circuits and mapping them onto quantum architectures.

\subsubsection{Quantum Point Addition}
As the most expensive composite arithmetic operation in Shor’s algorithm for breaking ECC, point addition is reviewed here in detail. Similar to our earlier content, we focus on the composite operation itself rather than its individual building blocks.



In Amento et al. \cite{amento2012efficient}, the point addition on projective coordinates for binary elliptic curves requires many ancilla qubits, as well as separate input and output qubits, and leaves garbage qubits after the computation. This approach has the advantage of avoiding the inversion for implementing point addition. The point addition of \cite{amento2012efficient} is implemented using only 13 multiplications without inversion. 
Subsequently, Roetteler et al. \cite{roetteler2017quantum} adopted an affine Weierstrass representation in a prime field and realized an in-place point-addition with addition, multiplication, and inversion.
The in-place point addition does not increase the number of qubits, even when multiple point additions are performed in Shor’s algorithm. 
From the perspective of quantum cryptanalysis of ECC using Shor’s algorithm, the in-place point addition provides a significant advantage.
In 2020, Banegas et al. \cite{banegas2020concrete} adopted affine coordinates to implement point addition over binary field.
Unlike projective coordinates, inversion is required, which consumes significantly more quantum resources than basic arithmetic operations. 

In 2021, Häner et al.~\cite{haner2020improved} firstly applied a windowing strategy for the in-place point addition over a prime field.
This reduces the number of required point additions from $2n$ to $(2n)/w$, where $w$ denotes the window size. 
Furthermore, Jang et al. \cite{jang2025new} proposed a low-depth out-of-place point addition and an improved in-place design, both in affine coordinates for binary elliptic curves.
The authors employed the windowing technique to reduce the total number of point additions.\\







More recent progress in these composite operations comes from the introduction of Look-up Table (LUT) based arithmetic techniques~\cite{gidney2019windowed}, which substantially reduce the quantum resource requirements, as discussed in Section~\ref{sec: windowed}.
\section{Look-up Table (LUT)-based Quantum Arithmetic}

\subsubsection{Windowed Arithmetic\label{sec: windowed}}
LUT-based Quantum Arithmetic, often referred to as windowed arithmetic, provides an alternative to computation-based approaches by reducing the number of quantum operations through the use of precomputed lookup tables. 
In 2019, this concept was firstly introduced to quantum computing by Gidney~\cite{gidney2019windowed}, who adapted the classical windowing technique to the quantum setting. Specifically, this windowed design optimizes modular multiplication and exponentiation using Look-up Table, significantly reducing the required number of quantum multiplications. Although it increases the amount of classical processing, the trade-off is highly favorable for fault-tolerant quantum architectures, where quantum operations are far more expensive than classical ones. 
Subsequent work by Luongo et al.~\cite{luongo2025optimizing} refined the initial windowed designs by optimizing the uncomputation procedures of memory lookups. Their techniques are particularly effective in settings that involve multiple consecutive table lookups, leading to further reductions in both Toffoli-Count and circuit depth.

\subsubsection{Windowed Modular Exponentiation}

In \cite{gidney2021factor,gidney2025factor}, the modular exponentiation subroutine in Shor’s algorithm, which originally requires $2n$ control qubits, is implemented using windowed quantum arithmetic. Instead of performing one modular multiplication per control qubit, each group of $w$ control qubits is used as the address of a quantum look-up table. Based on this address, the circuit loads a classically precomputed power and performs a single windowed modular multiplication. Consequently, the number of controlled modular multiplications is reduced from $n$ to $n/w$. Note that increasing the window size reduces the number of modular multiplications, but it also increases the cost of constructing and accessing the quantum look-up table. Therefore, an appropriate window size must be chosen to achieve a balanced trade-off.





\subsubsection{Windowed Point Addition}

Shor's algorithm uses $2n$ control qubits, each of which triggers a separate controlled point addition, leading to $2n$ point addition. 
In~\cite{haner2020improved}, these controls are grouped into windows of size~$w$. 
From the perspective of point addition, the inner arithmetic remains unchanged, but each window requires a quantum look-up.
Every $w$-bit register is used as an address of the quantum look-up table of precomputed points in superposition (i.e., $[1]P_2, [2]P_2, \ldots, [2^{w}-1]P_2$), and six look-ups are required per a single point addition.
This window technique effectively reduces the number of controlled point additions from $2n$ to $2n/w$.

\section{Applications toward Quantum Cryptanalysis}
In this section, we review applications of the quantum arithmetic circuits toward quantum public-key cryptanalysis, focusing on Rivest -- Shamir -- Adleman (RSA) cryptosystem and Elliptic Curve Cryptography (ECC).

\subsection{Rivest -- Shamir -- Adleman (RSA)}

In the work of Gidney and Ekerå \cite{gidney2021factor}, the modular exponentiation (which is the primary arithmetic operation in Shor’s algorithm for RSA), is optimized using three techniques: windowed arithmetic using quantum look-up table and the coset representation combined with oblivious carry runways \cite{gidney2019approximate}.
First, \textbf{windowed arithmetic} is applied both at the exponent level and inside each multiplication. 
At the exponent level, the standard square-and-multiply structure for modular exponentiation with windowing arithmetic. 
At the multiplication level, windowing allows multiple controlled additions to be merged into a single look-up addition.
\textbf{Coset representation} is used to implement modular addition as a cheaper non-modular addition with only a small approximation error. It effectively eliminates the need for comparison and multiple conditional subtractions in modular arithmetic.
For \textit{oblivious carry runways}, additions are partitioned across multiple registers that do not interfere with one another, allowing carry propagation to be localized within each segment and enabling substantial parallelism in the overall addition process.
As a result, these techniques reduce both the cost and the depth of modular multiplication, leading to a significantly more efficient realization of modular exponentiation for Shor’s algorithm.

Most recently, in Gidney’s 2025 work~\cite{luongo2025optimizing}, the controlled constant modular multiplication remains the core building block, and two major optimizations are introduced: The Chinese Remainder Theorem (CRT) and approximate residue arithmetic. Instead of performing modular multiplications over a large modulus, Gidney distributes the computation across many smaller moduli using CRT. This decomposition replaces large-modulus arithmetic with multiple small-modulus operations of significantly lower cost.
Approximate residue arithmetic keeps only the upper bits and discards the lower bits, thereby reducing the cost of implementation. The CRT structure makes this approximation viable because errors introduced by truncating lower bits in each small modulus remain locally bounded. Truncating low bits under a single large modulus would cause uncontrolled deviations, whereas truncating them within each small modulus produces only small local errors. As a result, the overall deviation remains small enough to preserve the periodicity required for Shor's algorithm. By discarding these low-order bits, the arithmetic becomes substantially cheaper, while maintaining sufficient accuracy for the period-finding procedure.

\subsection{Elliptic Curve Cryptography (ECC)}

\subsubsection{Binary ECC}

In 2012 and 2020, Amento et al. \cite{amento2012efficient} and Banegas et al. \cite{banegas2020concrete} presented quantum cryptanalysis of binary elliptic curves. The main difference between \cite{amento2012efficient} and \cite{banegas2020concrete} lies in the coordinate system adopted for implementing point addition quantum circuits. 

The authors of \cite{banegas2020concrete} adopted affine coordinates to implement point addition, defined as $P_3(x_3, y_3) = P_1(x_1, y_1) + P_2(x_2, y_2)$, where $x_3 = \lambda^2 + \lambda + x_1 + x_2 + a$, $y_3 = (x_2 + x_3)\lambda + x_3 + y_2$, and $\lambda = \frac{y_1 + y_2}{x_1 + x_2}$. To compute $\lambda$, the inversion of $(x_1 + x_2)$ is required, which consumes significantly more quantum resources than basic arithmetic operations such as addition, squaring, and multiplication. 
On the other hand, the authors of \cite{amento2012efficient} adopted projective coordinates, and the point addition is implemented without inversion, as mentioned earlier.

%
%
It is important to note that the point addition in \cite{banegas2020concrete} (on affine coordinates) computes the output result on the input (i.e., in-place) and initializes all the ancilla qubits used for intermediate results through reverse operations. 
In contrast, the point addition in \cite{amento2012efficient} (on projective coordinates) requires many ancilla qubits, as well as separate input and output qubits, and leaves garbage qubits after the computation.
Recall that Shor’s algorithm performs multiple point additions (i.e., not a stand-alone point addition). Thus, if the point addition in \cite{amento2012efficient} is adopted, it generates garbage qubits for every point addition. In contrast, the point addition in \cite{banegas2020concrete} does not increase the number of qubits, even when multiple point additions are performed in Shor’s algorithm.

In \cite{jang2025new}, Jang et al. (2025) significantly improved the performance trade-off metric, defined as the product of circuit depth and Qubit-Count.
Similar to \cite{banegas2020concrete}, they adopted affine coordinates for implementing point addition circuits.
For multiplication, they employed Jang et al.’s Toffoli-depth-one multiplication circuit \cite{jang2022optimized}, which is based on the Karatsuba algorithm.
They also presented a Fermat’s Little Theorem (FLT)–based inversion consisting of multiple multiplication and squaring operations.
Since their inversion leverages low-depth multiplication, it also achieves a low depth while demonstrating the best trade-off performance (i.e., the product of depth and Qubit-Count) compared with previous work \cite{banegas2020concrete}.
As a side note, \cite{banegas2020concrete} adopted Van Hoof’s Karatsuba multiplication \cite{van2019space} (optimized for Qubit-Count) and a greatest common divisor (GCD)-based inversion\footnote{GCD-based inversion is suitable for optimizing Qubit-Count, whereas FLT-based inversion is suitable for optimizing circuit depth.} for their point addition quantum circuits.

\subsubsection{Prime (Non-binary) ECC}
In 2017, Roetteler et al. \cite{roetteler2017quantum} presented the first practical quantum resource estimation for elliptic-curve cryptanalysis over prime fields. They adopted the affine coordinate system in Weierstrass form and introduced reversible arithmetic circuits for modular addition, Montgomery multiplication, and modular inversion used in point addition.
An in-place Takahashi adder was employed for addition throughout the circuit.
For multiplication, they designed a bitwise Montgomery multiplier consisting of controlled additions, shifts, and a single final correction of $\pm$ p.
This structure reduces circuit depth compared with add-and-double method but requires extra ancillas to store the parity information.
For inversion, they used Kaliski’s binary extended Euclidean algorithm \cite{kaliski2002montgomery}, which is composed of addition, subtraction, and cyclic shifts.

In 2021, Häner et al. \cite{haner2020improved} presented an improved quantum circuit architecture building upon \cite{roetteler2017quantum}.
They introduce three design trade-offs between width, T-gate, and depth to optimize circuit resources under different constraints.
For addition, the Takahashi, CDKM, and Draper adders were employed for the low-width, low-T, and low-depth configurations, respectively.
An efficient controlled-addition scheme was introduced, where only a subset of operations within the adder is controlled instead of applying control to the entire adder.
They also introduced an optimized Montgomery multiplication based on window arithmetic with quantum look-up reduction.
For each window, precomputed values were looked up according to the lower $k$ bits of the partial product (where $k$ denotes the window size) and used as reduction factors.
For inversion, they proposed a swap-based Kaliski algorithm that replaces certain controlled additions and shifts used in \cite{roetteler2017quantum} with conditional swaps.
At the level of Shor’s algorithm, they also applied the windowing technique, reducing the number of required point additions from $2n$ to $2n/k$.
Their results show lower quantum resource requirements in both depth and T-gate metrics compared with \cite{roetteler2017quantum}, indicating a notable improvement in overall circuit efficiency.



\section{Fault-Tolerant Circuit Mapping}

Quantum systems are innately fragile due to multiple sources of noise that affect circuit execution at various stages. 
Typical sources include \textit{decoherence}, \textit{crosstalk}, \textit{leakage}, and \textit{imperfect gate control}, 
all of which introduce errors that perturb the ideal quantum evolution. 
These errors can be modeled as a unitary disturbance \( U \) acting on a qubit, 
comprising different degrees of influence along the Pauli axes \( X, Y, Z \). 
Since \( Y = XZ \), any arbitrary quantum error can be represented as a combination of 
\textit{bit-flip} (\( X \)) and \textit{phase-flip} (\( Z \)) errors.
\[
\begin{aligned}
    U = \alpha I + \beta X + \gamma XZ + \delta Z.
\end{aligned}
\]
In quantum computing, mitigating the effects of noise requires quantum error correction (QEC), which introduces specialized coding schemes to detect and correct errors during quantum computation. Specifically, a single logical qubit is encoded into multiple physical qubits, and carefully designed parity (syndrome) measurements are performed to identify errors without collapsing the encoded quantum information. By repeatedly extracting this syndrome information and applying appropriate recovery operations, QEC suppresses the accumulation of both bit-flip and phase-flip errors, enabling reliable quantum computation over long circuit depths.

While QEC enables reliable quantum operation, it also incurs significant overheads in both qubit resources and execution time. Among quantum gates, the T gate is particularly costly to implement in a fault-tolerant manner because it cannot be implemented transversally in most stabilizer codes. Instead, T gate must be realized through state injection and magic-state distillation~\cite{Sales_Rodriguez_2025}. This distillation process consumes many ancilla qubits and additional logical cycles, and must be repeated whenever the error rate of the magic states exceeds a certain threshold. As a result, magic-state distillation becomes a major contributor to the overall time-space complexity of fault-tolerant quantum computing.

In brief, accurate runtime estimates~\cite{gidney2021factor,gidney2025factor} for fault-tolerant quantum circuits depend critically on the choice of error correcting code and the specific magic-state distillation protocols employed. 

\subsection{Quantum Error Correcting Codes}
Quantum error correction was initially inspired by the classical repetition codes, and this led to the development of codes like the Shor and Steane codes, which are the founding methodologies in stabilizer and CSS codes. However, these codes are not scalable in the process of continuous error detection and corrections. In current days, most experimental and architectural work revolves around 2D surface codes \cite{FowlerSurfaceCode}  and their variant, which offer high thresholds around the 1\% level and are well matched to planar hardware; a comparison of real codes implemented in hardware can be found in Table \ref{tab:hardware_comp}.

\begin{table*}[ht]
\centering
\caption{Comparison of Quantum Error-Correcting Code Families Implemented in Real Devices}
\label{tab:hardware_comp}
\resizebox{\textwidth}{!}{
\renewcommand{\arraystretch}{1.2}

\setlength{\tabcolsep}{3pt}
\begin{tabular}{p{2.5cm}| p{4cm}| p{3.2cm}| p{3cm}}
\toprule
\textbf{Code Type} 
& \textbf{Demonstrations} 
& \textbf{Physical Qubits} 
& \textbf{Tolerable Physical Error (\%)} 
\\
\midrule

{Surface Code} \cite{FowlerSurfaceCode} 
&  Google Quantum AI distance-5 over distance-3 surface code \cite{google2023suppressing};distance-7 code with lesser error rate on Willow \cite{google2025quantum} 
& 49-qubit distance-5 \cite{google2023suppressing};101-distance 7 \cite{google2025quantum}
& $2.914 \pm 0.016$ per cycle \cite{google2023suppressing}; $0.143\pm0.003$ per cycle \cite{google2025quantum}
\\ \hline

{Color Code} \cite{bombincolorcode}
& Improved Qubit scaling from surface code by color codes \cite{lacroix2025scaling}
& 37 Qubits distance-5 \cite{lacroix2025scaling}
& $1.20 \pm 0.02$ \% per cycle \cite{lacroix2025scaling};
\\ \hline

{GKP Code} \cite{gottesmangkp} &
High-quality GKP states generated in trapped ions \cite{fluhmann2019};
GKP error-correction cycles demonstrated in superconducting cavities \cite{campagne2020quantum}. &
1 Bosonic mode (encodes a qubit);
plus ancilla qubit for syndrome extraction &
Demonstrated operation at $\sim1\%$–$3\%$ squeezing-equivalent noise \\ \hline

{Cat Codes} \cite{catcode} &
Yale / AWS cavity-QED cat qubits achieving break-even quantum memory \cite{ofek2016extending};
Autonomous cat-code QEC via engineered dissipation \cite{lescanne2020exponential}. &
1 superconducting cavity + 1–2 ancillas &
Break-even lifetimes achieved at physical error $\sim0.5\%$–$1\%$ \\ \hline

{Concatenated Codes (5-qubit \cite{laflamme}, Steane \cite{steanecode})} &
Superconducting demonstrations of [[5,1,3]] code stabilizers \cite{reed2012realization}. &
5–7 qubits per logical block &
Logical error suppression observed for physical error rates $<1\%$ \\

\bottomrule
\end{tabular}}
\end{table*}
\vspace{-0.5cm}

\subsection{Runtime Estimations for Quantum Circuits}
The runtime estimation for fault-tolerant quantum circuits is obtained by computing the number of cycles required to execute the entire
logical circuit for a chosen error-correcting code. Here, we illustrate this approach using the surface code, which is a widely explored example.
We first choose an appropriate magic-state distillation scheme \cite{Litinski_2019,Bravyi_2012}, which is the major overhead in these estimations. Given this choice, we set a
target logical error rate and iteratively determine the required code distance
and the number of distillation rounds per layer. Using these distances, we then
compute the total number of physical qubits required for both data blocks and
magic-state factories. This is a standard methodology used in several works like \cite{amy2016estimating,di2020fault,cryptoeprint:2025/1832,GHEORGHIU2025107480}. A sample code for simulating this process is available online\footnote{\url{https://github.com/amitsaha2806/Surfacecode_Crypto}}. However, there has also been some work on the grounds of estimating costs of a quantum circuit through compiling circuits into lattice surgery instructions, taking a more compiler-based mapping approach to hardware tiles \cite{Leblondlatticecompilation}. The authors include a magic state storage scheduling variable, and consumption can make storage overhead dominate. In \cite{chemistryerrorcorrection}, the authors demonstrate full compilation of a chemistry circuit to lattice-surgery instructions and estimate the resource cost of the compiled circuit. In all resource estimation tasks, it is evident that distillation procedures consume the most amount of resources. To optimize this, there have been other works as well; \cite{leecolorcodemagic} proposes two new schemes for preparing high-fidelity magic states using 2D color codes \cite{bombincolorcode}. They compare the performance (overhead vs target error rate) with existing magic-state distillation methods for color codes, showing their versions use significantly fewer resources. Their source-code repository is available  online\footnote{\url{https://github.com/seokhyung-lee/msd-magic-state-prep-cycle-simulation}}. 

In recent times the resource estimation flow is aimed to be done through software that spans the algorithm-to-hardware stack. For example, Qualtran \cite{harrigan2024qualtran} (from Google Research) and Azure Quantum \cite{Azure_Quantum_Resource_Estimator} resource estimator enable architecture-independent modelling of quantum algorithms, automatically producing logical-level metrics such as T-gate count, logical depth, and magic-state requirements, which can subsequently be fed into cost models assuming a surface-code architecture.

\section{Conclusion}
This chapter examined the role of quantum arithmetic circuits in public-key cryptanalysis, focusing on the construction and optimization of fundamental arithmetic operations in Shor’s algorithm~\cite{Shor}. Specifically, techniques such as measurement-based uncomputation, windowed arithmetic, and conditionally clean ancilla methods provide promising pathways for reducing the quantum cost of arithmetic designs and have already inspired several new quantum implementations for breaking RSA and ECC. Moreover, we also highlighted the crucial role of quantum error correction in resource estimation.

Looking forward, further advances in design optimization and error correction will be important for assessing the feasibility of large-scale quantum cryptanalysis and for informing the design of quantum-resistant cryptographic standards.

\section*{Acknowledgments}
The authors sincerely thank Dr. Amit Saha for his valuable guidance, insightful feedback, and helpful discussions that contributed to the development of this chapter.

%
%
%
\bibliographystyle{splncs04}
\bibliography{main.bib}

@article{cheng2002quantum,
  title={Quantum full adder and subtractor},
  author={Cheng, Kai-Wen and Tseng, Chien-Cheng},
  journal={Electronics Letters},
  volume={38},
  number={22},
  pages={1343--1344},
  year={2002},
  publisher={IET}
}

@inproceedings{thapliyal2009design,
  title={Design of efficient reversible binary subtractors based on a new reversible gate},
  author={Thapliyal, Himanshu and Ranganathan, Nagarajan},
  booktitle={2009 IEEE computer society annual symposium on VLSI},
  pages={229--234},
  year={2009},
  organization={IEEE}
}

@incollection{thapliyal2016mapping,
  title={Mapping of subtractor and adder-subtractor circuits on reversible quantum gates},
  author={Thapliyal, Himanshu},
  booktitle={Transactions on Computational Science XXVII},
  pages={10--34},
  year={2016},
  publisher={Springer}
}

@misc{Gidney_Increment,
author = {Craig Gidney},
title = {Constructing Large Increment Gates},
year = {June 2015},
howpublished = {\url{https://algassert.com/circuits/2015/06/12/Constructing-Large-Increment-Gates.html}},
note = {Blog: Algorithmic Assertions}
}

@article{gidney2025classical,
  title={A Classical-Quantum Adder with Constant Workspace and Linear Gates},
  author={Gidney, Craig},
  journal={arXiv preprint arXiv:2507.23079},
  year={2025}
}

@article{haner2016factoring,
  title={Factoring using 2n+ 2 qubits with Toffoli based modular multiplication},
  author={H{\"a}ner, Thomas and Roetteler, Martin and Svore, Krysta M},
  journal={arXiv preprint arXiv:1611.07995},
  year={2016}
}

@article{gu2025resource,
  title={Resource analysis of Shor's elliptic curve algorithm with an improved quantum adder on a two-dimensional lattice},
  author={Gu, Quan and Ye, Han and Chen, Junjie and Ma, Xiongfeng},
  journal={arXiv preprint arXiv:2510.23212},
  year={2025}
}

@inproceedings{remaud2025ancilla,
  title={Ancilla-free quantum adder with sublinear depth},
  author={Remaud, Maxime and Vandaele, Vivien},
  booktitle={International Conference on Reversible Computation},
  pages={137--154},
  year={2025},
  organization={Springer}
}

@article{kim2025tree,
  title={Tree-based Quantum Carry-Save Adder},
  author={Kim, Hyunjun and Lim, Sejin and Jang, Kyungbae and Wang, Siyi and Baksi, Anubhab and Chattopadhyay, Anupam and Seo, Hwajeong},
  journal={Cryptology ePrint Archive},
  year={2025}
}

@article{Takahashi09,
author = {Takahashi, Yasuhiro and Tani, Seiichiro and Kunihiro, Noboru},
year = {2009},
month = {10},
pages = {},
title = {Quantum Addition Circuits and Unbounded Fan-Out},
volume = {10},
journal = {Quantum Information and Computation},
doi = {10.26421/QIC10.9-10-12}
}

@article{VBE,
author = {Vedral, Vlatko and Barenco, Adriano and Ekert, Artur},
year = {1995},
month = {11},
pages = {},
title = {Quantum Networks for Elementary Arithmetic Operations},
volume = {54},
journal = {Physical Review A},
doi = {10.1103/PhysRevA.54.147}
}

@misc{Cuccaro,
      title={A new quantum ripple-carry addition circuit}, 
      author={Steven A. Cuccaro and Thomas G. Draper and Samuel A. Kutin and David Petrie Moulton},
      year={2004},
      eprint={quant-ph/0410184},
      archivePrefix={arXiv},
      primaryClass={quant-ph}
}

@article{Draper,
author = {Draper, Thomas and Kutin, Samuel and Rains, Eric and Svore, Krysta},
year = {2004},
month = {07},
pages = {},
title = {A logarithmic-depth quantum carry-lookahead adder},
volume = {6},
journal = {Quantum Information and Computation},
doi = {10.26421/QIC6.4-5-4}
}

@article{gossett1998quantum,
      title={Quantum Carry-Save Arithmetic}, 
      author={Phil Gossett},
      year={1998},
      eprint={quant-ph/9808061},
      archivePrefix={arXiv},
      primaryClass={quant-ph}
}

@article{Takahashi05,
author = {Takahashi, Yasuhiro and Kunihiro, Noboru},
title = {A linear-size quantum circuit for addition with no ancillary qubits},
year = {2005},
issue_date = {September 2005},
publisher = {Rinton Press, Incorporated},
address = {Paramus, NJ},
volume = {5},
number = {6},
issn = {1533-7146},
journal = {Quantum Info. Comput.},
month = {sep},
pages = {440–448},
numpages = {9}
}

@article{Takahashi08,
author = {Takahashi, Yasuhiro and Kunihiro, Noboru},
year = {2008},
month = {07},
pages = {636-649},
title = {A fast quantum circuit for addition with few qubits},
volume = {8},
journal = {Quantum Information and Computation},
doi = {10.26421/QIC8.6-7-5}
}

@article{paper3_2021,
author = {S S, Gayathri and Kumar, R. and Samiappan, Dhanalakshmi and Kaushik, Brajesh Kumar and Haghparast, Majid},
year = {2021},
month = {08},
pages = {1-13},
title = {T-Count Optimized Wallace Tree Integer Multiplier for Quantum Computing},
volume = {60},
journal = {International Journal of Theoretical Physics},
doi = {10.1007/s10773-021-04864-3}
}

@article{paper4_2016,
author = {Wang, Feng and Luo, Mingxing and Li, Huiran and Qu, Zhiguo and Wang, Xiaojun},
year = {2016},
month = {02},
pages = {},
title = {Improved quantum ripple-carry addition circuit},
volume = {59},
journal = {Science China Information Sciences},
doi = {10.1007/s11432-015-5411-x}
}

@article{Gidney2018halvingcostof,
  doi = {10.22331/q-2018-06-18-74},
  url = {https://doi.org/10.22331/q-2018-06-18-74},
  title = {Halving the cost of quantum addition},
  author = {Gidney, Craig},
  journal = {{Quantum}},
  issn = {2521-327X},
  publisher = {{Verein zur F{\"{o}}rderung des Open Access Publizierens in den Quantenwissenschaften}},
  volume = {2},
  pages = {74},
  month = jun,
  year = {2018}
}

@ARTICLE{wang2023higher,
    journal = {Scientific Reports},
      title={A Higher Radix Architecture for Quantum Carry-lookahead Adder}, 
      author={Siyi Wang and Anubhab Baksi and Anupam Chattopadhyay},
      year={2023},
month = {09},
      volume = {13},
doi = {10.1038/s41598-023-41122-4}
}

@inproceedings{wang2023reducing,
    title={Reducing Depth of Quantum Adder using Ling Structure},
    author={Wang, Siyi and Chattopadhyay, Anupam},
    booktitle={VLSI-SoC},
    year={2023},
}

@inproceedings{wang2023efficient,
  title={Efficient Depth Optimization in Quantum Addition and Modular Arithmetic with Ling Structure},
  author={Wang, Siyi and Chattopadhyay, Anupam},
  booktitle={IFIP/IEEE International Conference on Very Large Scale Integration-System on a Chip},
  pages={73--89},
  year={2023},
  organization={Springer}
}

@article{wang2025optimal,
  title={Optimal toffoli-depth quantum adder},
  author={Wang, Siyi and Mondal, Ankit and Chattopadhyay, Anupam},
  journal={ACM Transactions on Quantum Computing},
  volume={6},
  number={3},
  pages={1--16},
  year={2025},
  publisher={ACM New York, NY}
}

@inproceedings{wang2024minimum,
  title={Minimum Depth Quantum Modular Addition Through Carry-Save Architecture},
  author={Wang, Siyi and Lim, Eugene and Li, Xiufan and Feng, Jerrie and Chattopadhyay, Anupam},
  booktitle={2024 IFIP/IEEE 32nd International Conference on Very Large Scale Integration (VLSI-SoC)},
  pages={1--6},
  year={2024},
  organization={IEEE}
}

@article{jang2023quantum,
  title={Quantum Binary Field Multiplication with Optimized Toffoli Depth and Extension to Quantum Inversion},
  author={Jang, Kyungbae and Kim, Wonwoong and Lim, Sejin and Kang, Yeajun and Yang, Yujin and Seo, Hwajeong},
  journal={Sensors},
  volume={23},
  number={6},
  pages={3156},
  year={2023},
  publisher={MDPI}
}

@article{zhang2025optimized,
  title={Optimized quantum folding Barrett reduction for quantum modular multipliers},
  author={Zhang, Jian and Cho, Seong-Min and Lee, Changyeol and Seo, Seung-Hyun},
  journal={Scientific Reports},
  volume={15},
  number={1},
  pages={22808},
  year={2025},
  publisher={Nature Publishing Group UK London}
}

@article{rosenberger1957simultaneous,
  title={Simultaneous carry adder},
  author={Rosenberger, Gerald B},
  journal={US Patent 2,966},
  volume={30},
  year={1957}
}

@misc{earle1967latched,
  title={Latched carry save adder circuit for multipliers},
  author={Earle, John G},
  year={1967},
  month=sep # "~5",
  publisher={Google Patents},
  note={US Patent 3,340,388}
}

@article{Sklansky,
author = {Sklansky, J.},
year = {1960},
month = {07},
pages = {226 - 231},
title = {Conditional-Sum Addition Logic},
volume = {EC-9},
journal = {Electronic Computers, IRE Transactions on},
doi = {10.1109/TEC.1960.5219822}
}

@article{li2022circuit,
  title={The circuit design and optimization of quantum multiplier and divider},
  author={Li, Hai-Sheng and Fan, Ping and Xia, Haiying and Long, Gui-Lu},
  journal={Science China Physics, Mechanics \& Astronomy},
  volume={65},
  number={6},
  pages={260311},
  year={2022},
  publisher={Springer}
}

@article{ananthalakshmi2017novel,
  title={A novel power efficient 0.64-GFlops fused 32-bit reversible floating point arithmetic unit architecture for digital signal processing applications},
  author={AnanthaLakshmi, AV and Sudha, Gnanou Florence},
  journal={Microprocessors and Microsystems},
  volume={51},
  pages={366--385},
  year={2017},
  publisher={Elsevier}
}

@article{thapliyal2019quantum,
  title={Quantum circuit designs of integer division optimizing T-count and T-depth},
  author={Thapliyal, Himanshu and Munoz-Coreas, Edgard and Varun, TSS and Humble, Travis S},
  journal={IEEE transactions on emerging topics in computing},
  volume={9},
  number={2},
  pages={1045--1056},
  year={2019},
  publisher={IEEE}
}

@article{gayathri2021t,
  title={T-count optimized quantum circuit designs for single-precision floating-point division},
  author={Gayathri, SS and Kumar, R and Dhanalakshmi, Samiappan and Dooly, Gerard and Duraibabu, Dinesh Babu},
  journal={Electronics},
  volume={10},
  number={6},
  pages={703},
  year={2021},
  publisher={MDPI}
}

@inproceedings{gayathri2022efficient,
  title={Efficient Floating-point Division Quantum Circuit using Newton-Raphson Division},
  author={Gayathri, SS and Kumar, R and Dhanalakshmi, Samiappan},
  booktitle={Journal of Physics: Conference Series},
  volume={2335},
  number={1},
  pages={012058},
  year={2022},
  organization={IOP Publishing}
}

@article{yuan2022novel,
  title={A novel fault-tolerant quantum divider and its simulation},
  author={Yuan, Suzhen and Gao, Shengwei and Wen, Chao and Wang, Yuchan and Qu, Hong and Wang, Yan},
  journal={Quantum Information Processing},
  volume={21},
  number={5},
  pages={182},
  year={2022},
  publisher={Springer}
}

@inproceedings{wang2024boosting,
  title={Boosting the efficiency of quantum divider through effective design space exploration},
  author={Wang, Siyi and Lim, Eugene and Chattopadhyay, Anupam},
  booktitle={2024 IEEE International Symposium on Circuits and Systems (ISCAS)},
  pages={1--5},
  year={2024},
  organization={IEEE}
}

@article{PhysRevA.109.052601,
  title = {Implementation of a quantum division circuit on noisy intermediate-scale quantum devices using dynamic circuits and approximate computing},
  author = {Sajadimanesh, Sohrab and Atoofian, Ehsan},
  journal = {Phys. Rev. A},
  volume = {109},
  issue = {5},
  pages = {052601},
  numpages = {12},
  year = {2024},
  month = {May},
  publisher = {American Physical Society},
  doi = {10.1103/PhysRevA.109.052601},
  url = {https://link.aps.org/doi/10.1103/PhysRevA.109.052601}
}

@article{orts2024quantum,
  title={Quantum circuit optimization of an integer divider},
  author={Orts, Francisco and Paulavi{\v{c}}ius, Remigijus and Filatovas, Ernestas},
  journal={Journal of Systems and Software},
  pages={112091},
  year={2024},
  publisher={Elsevier}
}

@phdthesis{goldschmidt1964applications,
  title={Applications of division by convergence},
  author={Goldschmidt, Robert E},
  year={1964},
  school={Massachusetts Institute of Technology}
}

@article{shaw1950arithmetic,
  title={Arithmetic operations in a binary computer},
  author={Shaw, Robert F},
  journal={Review of scientific instruments},
  volume={21},
  number={8},
  pages={687--693},
  year={1950},
  publisher={American Institute of Physics}
}

@article{harris1998srt,
  title={SRT division: Architectures, models, and implementations},
  author={Harris, D and Oberman, S and Horowitz, M},
  journal={Computer Systems Library, Standard University, Tech. Rep},
  year={1998},
  publisher={Citeseer}
}

@article{Bennett1973LogicalRO,
  title={Logical reversibility of computation},
  author={Charles H. Bennett},
  journal={Ibm Journal of Research and Development},
  year={1973},
  volume={17},
  pages={525-532},
  url={https://api.semanticscholar.org/CorpusID:14641793}
}

@article{cross2019validating,
  title={Validating quantum computers using randomized model circuits},
  author={Cross, Andrew W and Bishop, Lev S and Sheldon, Sarah and Nation, Paul D and Gambetta, Jay M},
  journal={Physical Review A},
  volume={100},
  number={3},
  pages={032328},
  year={2019},
  publisher={APS}
}

@inproceedings{10.1145/3672608.3707921,
author = {Hwang, Sungyoun and Seo, Hyoju and Kim, Yongtae},
title = {Can Less Accurate Be More Accurate? Surpassing Exact Multiplier with Approximate Design on NISQ Quantum Computers},
year = {2025},
isbn = {9798400706295},
publisher = {Association for Computing Machinery},
address = {New York, NY, USA},
url = {https://doi.org/10.1145/3672608.3707921},
doi = {10.1145/3672608.3707921},
booktitle = {Proceedings of the 40th ACM/SIGAPP Symposium on Applied Computing},
pages = {590–591},
numpages = {2},
location = {Catania International Airport, Catania, Italy},
series = {SAC '25}
}

@article{wardhani2024high,
  title={High-and half-degree quantum multiplication for post-quantum security evaluation},
  author={Wardhani, Rini Wisnu and Putranto, Dedy Septono Catur and Kim, Howon},
  journal={IEEE Access},
  volume={12},
  pages={8806--8821},
  year={2024},
  publisher={IEEE}
}

@article{kim2024toffoli,
  title={Toffoli gate count optimized space-efficient quantum circuit for binary field multiplication},
  author={Kim, Sunyeop and Kim, Insung and Kim, Seonggyeom and Hong, Seokhie},
  journal={Quantum Information Processing},
  volume={23},
  number={10},
  pages={330},
  year={2024},
  publisher={Springer}
}

@inproceedings{wang2025reducing,
  title={Reducing T-Depth and T-Count in Quantum Multiplication Using Compressor Primitives},
  author={Wang, Siyi and Dutta, Suman and Lee, Wei Jie Bryan and Feng, Jerrie and Fang, Xiang and Chattopadhyay, Anupam},
  booktitle={Proceedings of the Great Lakes Symposium on VLSI 2025},
  pages={35--40},
  year={2025}
}

@article{Toom-Cook-multi,
  title = {Quantum circuits for Toom-Cook multiplication},
  author = {Dutta, Srijit and Bhattacharjee, Debjyoti and Chattopadhyay, Anupam},
  journal = {Phys. Rev. A},
  volume = {98},
  issue = {1},
  pages = {012311},
  numpages = {6},
  year = {2018},
  month = {Jul},
  publisher = {American Physical Society},
  doi = {10.1103/PhysRevA.98.012311},
  url = {https://link.aps.org/doi/10.1103/PhysRevA.98.012311}
}

@article{larasati2021quantum,
  title={Quantum circuit design of Toom 3-way multiplication},
  author={Larasati, Harashta Tatimma and Awaludin, Asep Muhamad and Ji, Janghyun and Kim, Howon},
  journal={Applied Sciences},
  volume={11},
  number={9},
  pages={3752},
  year={2021},
  publisher={MDPI}
}

@article{putranto2023space,
  title={Space and time-efficient quantum multiplier in post quantum cryptography era},
  author={Putranto, Dedy Septono Catur and Wardhani, Rini Wisnu and Larasati, Harashta Tatimma and Kim, Howon},
  journal={IEEE Access},
  volume={11},
  pages={21848--21862},
  year={2023},
  publisher={IEEE}
}

@article{zalka1998fast,
  title={Fast versions of Shor's quantum factoring algorithm},
  author={Zalka, Christof},
  journal={arXiv preprint quant-ph/9806084},
  year={1998}
}

@article{parent2017improved,
  title={Improved reversible and quantum circuits for Karatsuba-based integer multiplication},
  author={Parent, Alex and Roetteler, Martin and Mosca, Michele},
  journal={arXiv preprint arXiv:1706.03419},
  year={2017}
}

@article{lin2014qlib,
  title={Qlib: Quantum module library},
  author={Lin, Chia-Chun and Chakrabarti, Amlan and Jha, Niraj K},
  journal={ACM Journal on Emerging Technologies in Computing Systems (JETC)},
  volume={11},
  number={1},
  pages={1--20},
  year={2014},
  publisher={ACM New York, NY, USA}
}

@article{kepley2015quantum,
  title={Quantum circuits for F \_ 2\^{} n F 2 n-multiplication with subquadratic gate count},
  author={Kepley, Shane and Steinwandt, Rainer},
  journal={Quantum Information Processing},
  volume={14},
  pages={2373--2386},
  year={2015},
  publisher={Springer}
}

@article{jayashree2016ancilla,
  title={Ancilla-input and garbage-output optimized design of a reversible quantum integer multiplier},
  author={Jayashree, HV and Thapliyal, Himanshu and Arabnia, Hamid R and Agrawal, Vinod Kumar},
  journal={The Journal of Supercomputing},
  volume={72},
  pages={1477--1493},
  year={2016},
  publisher={Springer}
}

@article{munoz2018quantum,
  title={Quantum circuit design of a T-count optimized integer multiplier},
  author={Mu{\~n}oz-Coreas, Edgard and Thapliyal, Himanshu},
  journal={IEEE Transactions on Computers},
  volume={68},
  number={5},
  pages={729--739},
  year={2018},
  publisher={IEEE}
}

@article{gidney2019asymptotically,
  title={Asymptotically efficient quantum Karatsuba multiplication},
  author={Gidney, Craig},
  journal={arXiv preprint arXiv:1904.07356},
  year={2019}
}

@inproceedings{sajadimanesh2022practical,
  title={Practical approximate quantum multipliers for NISQ devices},
  author={Sajadimanesh, Sohrab and Faye, Jean Paul Latyr and Atoofian, Ehsan},
  booktitle={Proceedings of the 19th ACM International Conference on Computing Frontiers},
  pages={121--130},
  year={2022}
}

@article{orts2023improving,
  title={Improving the number of t gates and their spread in integer multipliers on quantum computing},
  author={Orts, F and Filatovas, E and Ortega, G and SanJuan-Estrada, JF and Garz{\'o}n, EM},
  journal={Physical Review A},
  volume={107},
  number={4},
  pages={042621},
  year={2023},
  publisher={APS}
}

@article{nie2023quantum,
  title={Quantum circuit design for integer multiplication based on Sch{\"o}nhage-Strassen algorithm},
  author={Nie, Junhong and Zhu, Qinlin and Li, Meng and Sun, Xiaoming},
  journal={IEEE Transactions on Computer-Aided Design of Integrated Circuits and Systems},
  year={2023},
  publisher={IEEE}
}

@inproceedings{karatsuba1962multiplication,
  title={Multiplication of many-digital numbers by automatic computers},
  author={Karatsuba, Anatolii Alekseevich and Ofman, Yu P},
  booktitle={Doklady Akademii Nauk},
  volume={145},
  number={2},
  pages={293--294},
  year={1962},
  organization={Russian Academy of Sciences}
}

@article{wallace1964suggestion,
  title={A suggestion for a fast multiplier},
  author={Wallace, Christopher S},
  journal={IEEE Transactions on electronic Computers},
  number={1},
  pages={14--17},
  year={1964},
  publisher={IEEE}
}

@article{cook1969minimum,
  title={On the minimum computation time of functions},
  author={Cook, Stephen A and Aanderaa, St{\aa}l O},
  journal={Transactions of the American Mathematical Society},
  volume={142},
  pages={291--314},
  year={1969},
  publisher={JSTOR}
}

@phdthesis{wang2025innovative,
  title={Innovative design and optimization of arithmetic circuits in quantum computing},
  author={Wang, Siyi},
  year={2025},
  school={Nanyang Technological University}
}

@article{wang2025comprehensive,
  title={A comprehensive study of quantum arithmetic circuits},
  author={Wang, Siyi and Li, Xiufan and Lee, Wei Jie Bryan and Deb, Suman and Lim, Eugene and Chattopadhyay, Anupam},
  journal={Philosophical Transactions A},
  volume={383},
  number={2288},
  pages={20230392},
  year={2025},
  publisher={The Royal Society}
}

@article{gidney2019windowed,
  title={Windowed quantum arithmetic},
  author={Gidney, Craig},
  journal={arXiv preprint arXiv:1905.07682},
  year={2019}
}

@inproceedings{luongo2025optimizing,
  title={Optimizing windowed arithmetic for quantum attacks against RSA-2048},
  author={Luongo, Alessandro and Narasimhachar, Varun and Sireesh, Adithya},
  booktitle={2025 62nd ACM/IEEE Design Automation Conference (DAC)},
  pages={1--7},
  year={2025},
  organization={IEEE}
}

@article{gidney2021factor,
  title={How to factor 2048 bit RSA integers in 8 hours using 20 million noisy qubits},
  author={Gidney, Craig and Eker{\aa}, Martin},
  journal={Quantum},
  volume={5},
  pages={433},
  year={2021},
  publisher={Verein zur F{\"o}rderung des Open Access Publizierens in den Quantenwissenschaften}
}

@inproceedings{haner2020improved,
  title={Improved quantum circuits for elliptic curve discrete logarithms},
  author={H{\"a}ner, Thomas and Jaques, Samuel and Naehrig, Michael and Roetteler, Martin and Soeken, Mathias},
  booktitle={International conference on post-quantum cryptography},
  pages={425--444},
  year={2020},
  organization={Springer}
}

@article{beckman1996efficient,
  title={Efficient networks for quantum factoring},
  author={Beckman, David and Chari, Amalavoyal N and Devabhaktuni, Srikrishna and Preskill, John},
  journal={Physical Review A},
  volume={54},
  number={2},
  pages={1034},
  year={1996},
  publisher={APS}
}

@article{van2005fast,
  title={Fast quantum modular exponentiation},
  author={Van Meter, Rodney and Itoh, Kohei M},
  journal={Physical Review A},
  volume={71},
  number={5},
  pages={052320},
  year={2005},
  publisher={APS}
}

@article{markov2012constant,
  title={Constant-optimized quantum circuits for modular multiplication and exponentiation},
  author={Markov, Igor L and Saeedi, Mehdi},
  journal={arXiv preprint arXiv:1202.6614},
  year={2012}
}

@article{haner2018optimizing,
  title={Optimizing quantum circuits for arithmetic},
  author={H{\"a}ner, Thomas and Roetteler, Martin and Svore, Krysta M},
  journal={arXiv preprint arXiv:1805.12445},
  year={2018}
}

@inproceedings{das2019optimizing,
  title={Optimizing Quantum Circuits for Modular Exponentiation},
  author={Das, Rakesh and Chattopadhyay, Anupam and Rahaman, Hafizur},
  booktitle={2019 32nd International Conference on VLSI Design and 2019 18th International Conference on Embedded Systems (VLSID)},
  pages={407--412},
  year={2019},
  organization={IEEE}
}

@article{viola2001constructing,
  title={Constructing qubits in physical systems},
  author={Viola, Lorenza and Knill, Emanuel and Laflamme, Raymond},
  journal={Journal of Physics A: Mathematical and General},
  volume={34},
  number={35},
  pages={7067},
  year={2001},
  publisher={IOP Publishing}
}

@article{shaw2008encoding,
  title={Encoding one logical qubit into six physical qubits},
  author={Shaw, Bilal and Wilde, Mark M and Oreshkov, Ognyan and Kremsky, Isaac and Lidar, Daniel A},
  journal={Physical Review A—Atomic, Molecular, and Optical Physics},
  volume={78},
  number={1},
  pages={012337},
  year={2008},
  publisher={APS}
}

@inproceedings{luongo2025measurement,
  title={Measurement-based uncomputation of quantum circuits for modular arithmetic},
  author={Luongo, Alessandro and Miti, Antonio Michele and Narasimhachar, Varun and Sireesh, Adithya},
  booktitle={2025 62nd ACM/IEEE Design Automation Conference (DAC)},
  pages={1--7},
  year={2025},
  organization={IEEE}
}

@article{bravyi2005universal,
  title={Universal quantum computation with ideal Clifford gates and noisy ancillas},
  author={Bravyi, Sergey and Kitaev, Alexei},
  journal={Physical Review A—Atomic, Molecular, and Optical Physics},
  volume={71},
  number={2},
  pages={022316},
  year={2005},
  publisher={APS}
}

@article{kaliski2002montgomery,
  title={The Montgomery inverse and its applications},
  author={Kaliski, Burton S},
  journal={IEEE transactions on computers},
  volume={44},
  number={8},
  pages={1064--1065},
  year={2002},
  publisher={IEEE}
}

@article{gidney2019approximate,
  title={Approximate encoded permutations and piecewise quantum adders},
  author={Gidney, Craig},
  journal={arXiv preprint arXiv:1905.08488},
  year={2019}
}

@article{kornerup2021tight,
  title={Tight bounds on the spooky pebble game: Recycling qubits with measurements},
  author={Kornerup, Niels and Sadun, Jonathan and Soloveichik, David},
  journal={arXiv preprint arXiv:2110.08973},
  year={2021}
}

@article{nie2024quantum,
  title={Quantum circuit for multi-qubit toffoli gate with optimal resource},
  author={Nie, Junhong and Zi, Wei and Sun, Xiaoming},
  journal={arXiv preprint arXiv:2402.05053},
  year={2024}
}

@article{khattar2024rise,
  title={Rise of conditionally clean ancillae for optimizing quantum circuits},
  author={Khattar, Tanuj and Gidney, Craig},
  journal={arXiv preprint arXiv:2407.17966},
  year={2024},
  publisher={Jul}
}

@article{PhysRevA.111.052611,
  title = {Exact space-depth trade-offs in multicontrolled Toffoli decomposition},
  author = {Dutta, Suman and Wang, Siyi and Baksi, Anubhab and Chattopadhyay, Anupam and Maitra, Subhamoy},
  journal = {Phys. Rev. A},
  volume = {111},
  issue = {5},
  pages = {052611},
  numpages = {12},
  year = {2025},
  month = {May},
  publisher = {American Physical Society},
  doi = {10.1103/PhysRevA.111.052611},
  url = {https://link-aps-org.remotexs.ntu.edu.sg/doi/10.1103/PhysRevA.111.052611}
}

@article{Shor,
author = {Shor, Peter},
year = {1996},
month = {10},
pages = {},
title = {Algorithms for Quantum Computation: Discrete Logarithms and Factoring},
journal = {Proceedings of 35th Annual Symposium on Foundations of Computer Science},
doi = {10.1109/SFCS.1994.365700}
}

@article{gidney2025factor,
  title={How to factor 2048 bit RSA integers with less than a million noisy qubits},
  author={Gidney, Craig},
  journal={arXiv preprint arXiv:2505.15917},
  year={2025}
}

@inproceedings{grover1996fast,
  title={A fast quantum mechanical algorithm for database search},
  author={Grover, Lov K},
  booktitle={Proceedings of the twenty-eighth annual ACM symposium on Theory of computing},
  pages={212--219},
  year={1996}
}

@article{rivest1978method,
  title={A method for obtaining digital signatures and public-key cryptosystems},
  author={Rivest, Ronald L and Shamir, Adi and Adleman, Leonard},
  journal={Communications of the ACM},
  volume={21},
  number={2},
  pages={120--126},
  year={1978},
  publisher={ACM New York, NY, USA}
}

@article{banegas2020concrete,
  title={Concrete quantum cryptanalysis of binary elliptic curves},
  author={Banegas, Gustavo and Bernstein, Daniel J and Van Hoof, Iggy and Lange, Tanja},
  journal={Cryptology ePrint Archive},
  year={2020}
}

@article{amento2012efficient,
  title={Efficient quantum circuits for binary elliptic curve arithmetic: reducing T-gate complexity},
  author={Amento, Brittanney and Steinwandt, Rainer and Roetteler, Martin},
  journal={arXiv preprint arXiv:1209.6348},
  year={2012}
}

@article{jang2025new,
  title={New Quantum Cryptanalysis of Binary Elliptic Curves},
  author={Jang, Kyungbae and Srivastava, Vikas and Baksi, Anubhab and Sarkar, Santanu and Seo, Hwajeong},
  journal={IACR Transactions on Cryptographic Hardware and Embedded Systems},
  volume={2025},
  number={2},
  pages={781--804},
  year={2025}
}

@inproceedings{jang2022optimized,
  title={Optimized implementation of quantum binary field multiplication with toffoli depth one},
  author={Jang, Kyungbae and Kim, Wonwoong and Lim, Sejin and Kang, Yeajun and Yang, Yujin and Seo, Hwajeong},
  booktitle={International Conference on Information Security Applications},
  pages={251--264},
  year={2022},
  organization={Springer}
}

@article{van2019space,
  title={Space-efficient quantum multiplication of polynomials for binary finite fields with sub-quadratic Toffoli gate count},
  author={Van Hoof, Iggy},
  journal={arXiv preprint arXiv:1910.02849},
  year={2019}
}

@inproceedings{roetteler2017quantum,
  title={Quantum resource estimates for computing elliptic curve discrete logarithms},
  author={Roetteler, Martin and Naehrig, Michael and Svore, Krysta M and Lauter, Kristin},
  booktitle={International Conference on the Theory and Application of Cryptology and Information Security},
  pages={241--270},
  year={2017},
  organization={Springer}
}

@article{Sales_Rodriguez_2025,
   title={Experimental demonstration of logical magic state distillation},
   volume={645},
   ISSN={1476-4687},
   url={http://dx.doi.org/10.1038/s41586-025-09367-3},
   DOI={10.1038/s41586-025-09367-3},
   number={8081},
   journal={Nature},
   publisher={Springer Science and Business Media LLC},
   author={Sales Rodriguez, Pedro and Robinson, John M. and Jepsen, Paul Niklas and He, Zhiyang and Duckering, Casey and Zhao, Chen and Wu, Kai-Hsin and Campo, Joseph and Bagnall, Kevin and Kwon, Minho and Karolyshyn, Thomas and Weinberg, Phillip and Cain, Madelyn and Evered, Simon J. and Geim, Alexandra A. and Kalinowski, Marcin and Li, Sophie H. and Manovitz, Tom and Amato-Grill, Jesse and Basham, James I. and Bernstein, Liane and Braverman, Boris and Bylinskii, Alexei and Choukri, Adam and DeAngelo, Robert J. and Fang, Fang and Fieweger, Connor and Frederick, Paige and Haines, David and Hamdan, Majd and Hammett, Julian and Hsu, Ning and Hu, Ming-Guang and Huber, Florian and Jia, Ningyuan and Kedar, Dhruv and Kornjača, Milan and Liu, Fangli and Long, John and Lopatin, Jonathan and Lopes, Pedro L. S. and Luo, Xiu-Zhe and Macrì, Tommaso and Marković, Ognjen and Martínez-Martínez, Luis A. and Meng, Xianmei and Ostermann, Stefan and Ostroumov, Evgeny and Paquette, David and Qiang, Zexuan and Shofman, Vadim and Singh, Anshuman and Singh, Manuj and Sinha, Nandan and Thoreen, Henry and Wan, Noel and Wang, Yiping and Waxman-Lenz, Daniel and Wong, Tak and Wurtz, Jonathan and Zhdanov, Andrii and Zheng, Laurent and Greiner, Markus and Keesling, Alexander and Gemelke, Nathan and Vuletić, Vladan and Kitagawa, Takuya and Wang, Sheng-Tao and Bluvstein, Dolev and Lukin, Mikhail D. and Lukin, Alexander and Zhou, Hengyun and Cantú, Sergio H.},
   year={2025},
   month=jul, pages={620–625} }

@inproceedings{amy2016estimating,
  title={Estimating the cost of generic quantum pre-image attacks on SHA-2 and SHA-3},
  author={Amy, Matthew and Di Matteo, Olivia and Gheorghiu, Vlad and Mosca, Michele and Parent, Alex and Schanck, John},
  booktitle={International Conference on Selected Areas in Cryptography},
  pages={317--337},
  year={2016},
  organization={Springer}
}

@misc{cryptoeprint:2025/1832,
      author = {Anik Basu Bhaumik and Suman Dutta and Siyi Wang and Anubhab Baksi and Kyungbae Jang and Amit Saha and Hwajeong Seo and Anupam Chattopadhyay},
      title = {Can Quantum Break {ZUC}? Only with a Million Qubits and a Billion Years to Spare},
      howpublished = {Cryptology {ePrint} Archive, Paper 2025/1832},
      year = {2025},
      url = {https://eprint.iacr.org/2025/1832}
}

@article{di2020fault,
  title={Fault-tolerant resource estimation of quantum random-access memories},
  author={Di Matteo, Olivia and Gheorghiu, Vlad and Mosca, Michele},
  journal={IEEE Transactions on Quantum Engineering},
  volume={1},
  pages={1--13},
  year={2020},
  publisher={IEEE}
}

@article{Leblondlatticecompilation,
author = {Leblond, Tyler and Dean, Christopher and Watkins, George and Bennink, Ryan},
title = {Realistic Cost to Execute Practical Quantum Circuits using Direct Clifford+T Lattice Surgery Compilation},
year = {2024},
issue_date = {December 2024},
publisher = {Association for Computing Machinery},
address = {New York, NY, USA},
volume = {5},
number = {4},
url = {https://doi-org.remotexs.ntu.edu.sg/10.1145/3689826},
doi = {10.1145/3689826},
journal = {ACM Transactions on Quantum Computing},
month = oct,
articleno = {25},
numpages = {28}
}

@article{chemistryerrorcorrection,
  title = {Compilation of a simple chemistry application to quantum error correction primitives},
  author = {Blunt, Nick S. and Geh\'er, Gy\"orgy P. and Moylett, Alexandra E.},
  journal = {Phys. Rev. Res.},
  volume = {6},
  issue = {1},
  pages = {013325},
  numpages = {20},
  year = {2024},
  month = {Mar},
  publisher = {American Physical Society},
  doi = {10.1103/PhysRevResearch.6.013325},
  url = {https://link.aps.org/doi/10.1103/PhysRevResearch.6.013325}
}

@article{leecolorcodemagic,
  title = {Low-Overhead Magic State Distillation with Color Codes},
  author = {Lee, Seok-Hyung and Thomsen, Felix and Fazio, Nicholas and Brown, Benjamin J. and Bartlett, Stephen D.},
  journal = {PRX Quantum},
  volume = {6},
  issue = {3},
  pages = {030317},
  numpages = {50},
  year = {2025},
  month = {Jul},
  publisher = {American Physical Society},
  doi = {10.1103/ch5r-cnfq},
  url = {https://link.aps.org/doi/10.1103/ch5r-cnfq}
}

@article{bombincolorcode,
  title = {Topological Quantum Distillation},
  author = {Bombin, H. and Martin-Delgado, M. A.},
  journal = {Phys. Rev. Lett.},
  volume = {97},
  issue = {18},
  pages = {180501},
  numpages = {4},
  year = {2006},
  month = {Oct},
  publisher = {American Physical Society},
  doi = {10.1103/PhysRevLett.97.180501},
  url = {https://link.aps.org/doi/10.1103/PhysRevLett.97.180501}
}

@article{Litinski_2019,
   title={Magic State Distillation: Not as Costly as You Think},
   volume={3},
   ISSN={2521-327X},
   url={http://dx.doi.org/10.22331/q-2019-12-02-205},
   DOI={10.22331/q-2019-12-02-205},
   journal={Quantum},
   publisher={Verein zur Forderung des Open Access Publizierens in den Quantenwissenschaften},
   author={Litinski, Daniel},
   year={2019},
   month=dec, pages={205} }

@article{Bravyi_2012,
   title={Magic-state distillation with low overhead},
   volume={86},
   ISSN={1094-1622},
   url={http://dx.doi.org/10.1103/PhysRevA.86.052329},
   DOI={10.1103/physreva.86.052329},
   number={5},
   journal={Physical Review A},
   publisher={American Physical Society (APS)},
   author={Bravyi, Sergey and Haah, Jeongwan},
   year={2012},
   month=nov }

@article{GHEORGHIU2025107480,
title = {Quantum resource estimation for large scale quantum algorithms},
journal = {Future Generation Computer Systems},
volume = {162},
pages = {107480},
year = {2025},
issn = {0167-739X},
doi = {https://doi.org/10.1016/j.future.2024.107480},
url = {https://www.sciencedirect.com/science/article/pii/S0167739X24004308},
author = {Vlad Gheorghiu and Michele Mosca}
}

@misc{harrigan2024qualtran,
    title={Expressing and Analyzing Quantum Algorithms with Qualtran},
    author={Matthew P. Harrigan and Tanuj Khattar
        and Charles Yuan and Anurudh Peduri and Noureldin Yosri
        and Fionn D. Malone and Ryan Babbush and Nicholas C. Rubin},
    year={2024},
    eprint={2409.04643},
    archivePrefix={arXiv},
    primaryClass={quant-ph},
    doi={10.48550/arXiv.2409.04643},
    url={https://arxiv.org/abs/2409.04643},
}

@inproceedings{wang2024quantumbook,
  title={Quantum Carry-Save Modular Addition with Optimized Depth and Resource Utilization},
  author={Wang, Siyi and Lim, Eugene and Li, Xiufan and Feng, Jerrie and Chattopadhyay, Anupam},
  booktitle={IFIP/IEEE International Conference on Very Large Scale Integration-System on a Chip},
  pages={17--32},
  year={2024},
  organization={Springer}
}

@inproceedings{Azure_Quantum_Resource_Estimator,
   author = {van Dam, Wim and Mykhailova, Mariia and Soeken, Mathias},
   title = {{Using Azure Quantum Resource Estimator for Assessing Performance of Fault Tolerant Quantum Computation}},
   year = {2023},
   isbn = {9798400707858},
   publisher = {Association for Computing Machinery},
   address = {New York, NY, USA},
   url = {https://doi.org/10.1145/3624062.3624211},
   doi = {10.1145/3624062.3624211},
   booktitle = {Proceedings of the SC '23 Workshops of The International Conference on High Performance Computing, Network, Storage, and Analysis},
   pages = {1414–1419},
   numpages = {6},
   series = {SC-W '23} }

@article{FowlerSurfaceCode,
  title = {Surface codes: Towards practical large-scale quantum computation},
  author = {Fowler, Austin G. and Mariantoni, Matteo and Martinis, John M. and Cleland, Andrew N.},
  journal = {Phys. Rev. A},
  volume = {86},
  issue = {3},
  pages = {032324},
  numpages = {48},
  year = {2012},
  month = {Sep},
  publisher = {American Physical Society},
  doi = {10.1103/PhysRevA.86.032324},
  url = {https://link.aps.org/doi/10.1103/PhysRevA.86.032324}
}

@article{google2025quantum,
  title={Quantum error correction below the surface code threshold},
  journal={Nature},
  volume={638},
  number={8052},
  pages={920--926},
  year={2025},
  publisher={Nature Publishing Group UK London}
}

@article{google2023suppressing,
  title={Suppressing quantum errors by scaling a surface code logical qubit},
  journal={Nature},
  volume={614},
  number={7949},
  pages={676--681},
  year={2023},
  publisher={Nature Publishing Group UK London}
}

@article{lacroix2025scaling,
  title={Scaling and logic in the color code on a superconducting quantum processor},
  author={Lacroix, Nathan and Bourassa, Alexandre and Heras, Francisco JH and Zhang, Lei M and Bausch, Johannes and Senior, Andrew W and Edlich, Thomas and Shutty, Noah and Sivak, Volodymyr and Bengtsson, Andreas and others},
  journal={Nature},
  pages={1--3},
  year={2025},
  publisher={Nature Publishing Group UK London}
}

@article{fluhmann2019,
  title={Encoding a qubit in a trapped-ion mechanical oscillator},
  author={Fl{\"u}hmann, Christa and Nguyen, Thanh Long and Marinelli, Matteo and Negnevitsky, Vlad and Mehta, Karan and Home, JP},
  journal={Nature},
  volume={566},
  number={7745},
  pages={513--517},
  year={2019},
  publisher={Nature Publishing Group UK London}
}

@article{campagne2020quantum,
  title={Quantum error correction of a qubit encoded in grid states of an oscillator},
  author={Campagne-Ibarcq, Philippe and Eickbusch, Alec and Touzard, Steven and Zalys-Geller, Evan and Frattini, Nicholas E and Sivak, Volodymyr V and Reinhold, Philip and Puri, Shruti and Shankar, Shyam and Schoelkopf, Robert J and others},
  journal={Nature},
  volume={584},
  number={7821},
  pages={368--372},
  year={2020},
  publisher={Nature Publishing Group UK London}
}

@article{ofek2016extending,
  title={Extending the lifetime of a quantum bit with error correction in superconducting circuits},
  author={Ofek, Nissim and Petrenko, Andrei and Heeres, Reinier and Reinhold, Philip and Leghtas, Zaki and Vlastakis, Brian and Liu, Yehan and Frunzio, Luigi and Girvin, Steven M and Jiang, Liang and others},
  journal={Nature},
  volume={536},
  number={7617},
  pages={441--445},
  year={2016},
  publisher={Nature Publishing Group UK London}
}

@article{lescanne2020exponential,
  title={Exponential suppression of bit-flips in a qubit encoded in an oscillator},
  author={Lescanne, Rapha{\"e}l and Villiers, Marius and Peronnin, Th{\'e}au and Sarlette, Alain and Delbecq, Matthieu and Huard, Benjamin and Kontos, Takis and Mirrahimi, Mazyar and Leghtas, Zaki},
  journal={Nature Physics},
  volume={16},
  number={5},
  pages={509--513},
  year={2020},
  publisher={Nature Publishing Group UK London}
}

@article{reed2012realization,
  title={Realization of three-qubit quantum error correction with superconducting circuits},
  author={Reed, Matthew D and DiCarlo, Leonardo and Nigg, Simon E and Sun, Luyan and Frunzio, Luigi and Girvin, Steven M and Schoelkopf, Robert J},
  journal={Nature},
  volume={482},
  number={7385},
  pages={382--385},
  year={2012},
  publisher={Nature Publishing Group UK London}
}

@article{gottesmangkp,
  title = {Encoding a qubit in an oscillator},
  author = {Gottesman, Daniel and Kitaev, Alexei and Preskill, John},
  journal = {Phys. Rev. A},
  volume = {64},
  issue = {1},
  pages = {012310},
  numpages = {21},
  year = {2001},
  month = {Jun},
  publisher = {American Physical Society},
  doi = {10.1103/PhysRevA.64.012310},
  url = {https://link.aps.org/doi/10.1103/PhysRevA.64.012310}
}

@article{catcode,
  title = {Quantum error correction using squeezed Schr\"odinger cat states},
  author = {Schlegel, David S. and Minganti, Fabrizio and Savona, Vincenzo},
  journal = {Phys. Rev. A},
  volume = {106},
  issue = {2},
  pages = {022431},
  numpages = {17},
  year = {2022},
  month = {Aug},
  publisher = {American Physical Society},
  doi = {10.1103/PhysRevA.106.022431},
  url = {https://link.aps.org/doi/10.1103/PhysRevA.106.022431}
}

@article{laflamme,
  title = {Perfect Quantum Error Correcting Code},
  author = {Laflamme, Raymond and Miquel, Cesar and Paz, Juan Pablo and Zurek, Wojciech Hubert},
  journal = {Phys. Rev. Lett.},
  volume = {77},
  issue = {1},
  pages = {198--201},
  numpages = {0},
  year = {1996},
  month = {Jul},
  publisher = {American Physical Society},
  doi = {10.1103/PhysRevLett.77.198},
  url = {https://link.aps.org/doi/10.1103/PhysRevLett.77.198}
}

@article{steanecode,
  title = {Error Correcting Codes in Quantum Theory},
  author = {Steane, A. M.},
  journal = {Phys. Rev. Lett.},
  volume = {77},
  issue = {5},
  pages = {793--797},
  numpages = {0},
  year = {1996},
  month = {Jul},
  publisher = {American Physical Society},
  doi = {10.1103/PhysRevLett.77.793},
  url = {https://link.aps.org/doi/10.1103/PhysRevLett.77.793}
}
\end{document}